\documentclass{article}

\usepackage{arxivsimplified}

\usepackage[utf8]{inputenc} 
\usepackage[T1]{fontenc}    
\usepackage{hyperref}       
\usepackage{url}            
\usepackage{booktabs}       
\usepackage{amsfonts}       
\usepackage{nicefrac}       
\usepackage{microtype}      
\usepackage{lipsum}

\usepackage{graphicx}
\usepackage{cite}
\usepackage[dvipsnames]{xcolor}
\usepackage{amsmath}
\usepackage{blindtext}
\usepackage{amsfonts}
\usepackage{multirow}
\usepackage{array}
\usepackage{mathtools}
\usepackage{listings}
\usepackage{xcolor}
\definecolor{codegreen}{rgb}{0,0.6,0}
\definecolor{codegray}{rgb}{0.5,0.5,0.5}
\definecolor{codepurple}{rgb}{0.58,0,0.82}
\definecolor{backcolour}{rgb}{0.95,0.95,0.92}
 
\lstdefinestyle{mystyle}{
    backgroundcolor=\color{backcolour},   
    commentstyle=\color{codegreen},
    keywordstyle=\color{magenta},
    numberstyle=\tiny\color{codegray},
    stringstyle=\color{codepurple},
    basicstyle=\ttfamily\footnotesize,
    breakatwhitespace=false,         
    breaklines=true,                 
    captionpos=b,                    
    keepspaces=true,                 
    numbers=left,                    
    numbersep=5pt,                  
    showspaces=false,                
    showstringspaces=false,
    showtabs=false,                  
    tabsize=2
}
 
\usepackage{scalerel,stackengine}
\stackMath
\newcommand\reallywidehat[1]{%
\savestack{\tmpbox}{\stretchto{%
  \scaleto{%
    \scalerel*[\widthof{\ensuremath{#1}}]{\kern-.6pt\bigwedge\kern-.6pt}%
    {\rule[-\textheight/2]{1ex}{\textheight}}
  }{\textheight}%
}{0.5ex}}%
\stackon[1pt]{#1}{\tmpbox}%
}

\title{A priori analysis on deep learning of subgrid-scale parameterizations for Kraichnan turbulence}

\author{
  Suraj Pawar  \\
  School of Mechanical \& Aerospace Engineering,\\
  Oklahoma State University, \\
  Stillwater, Oklahoma - 74078, USA.\\
  \texttt{supawar@okstate.edu} \\
   \And
 Omer San \\
 School of Mechanical \& Aerospace Engineering,\\
  Oklahoma State University, \\
  Stillwater, Oklahoma - 74078, USA.\\
  \texttt{osan@okstate.edu} \\
  \And
  Adil Rasheed  \\
  Department of Engineering Cybernetics,\\
  Norwegian University of Science and Technology,\\
  N-7465, Trondheim, Norway.\\
  \texttt{adil.rasheed@ntnu.no}\\
  \And
  Prakash Vedula \\
  School of Aerospace \& Mechanical Engineering,\\
  The University of Oklahoma, \\
  Norman, Oklahoma - 73019, USA.\\
  \texttt{pvedula@ou.edu}\\
}

\begin{document}
\maketitle

\begin{abstract}
In the present study, we investigate different data-driven parameterizations for large eddy simulation of two-dimensional turbulence in the \emph{a priori} settings. These models utilize resolved  flow field variables on the coarser grid to estimate the subgrid-scale stresses. We use data-driven closure models based on localized learning that employs multilayer feedforward artificial neural network (ANN) with point-to-point mapping and neighboring stencil data mapping, and convolutional neural network (CNN) fed by data snapshots of the whole domain. The performance of these data-driven closure models is measured through a probability density function and is compared with the dynamic Smagorinksy model (DSM). The quantitative performance is evaluated using the cross-correlation coefficient between the true and predicted stresses. We analyze different frameworks in terms of the amount of training data, selection of input and output features, their characteristics in modeling with accuracy, and training and deployment computational time. We also demonstrate computational gain that can be achieved using the intelligent eddy viscosity model that learns eddy viscosity computed by the DSM instead of subgrid-scale stresses. We detail the hyperparameters optimization of these models using the grid search algorithm.
\end{abstract}


\section{Introduction}
\label{sec:introduction}
Direct numerical simulation (DNS) of complex fluid flows encountered in many engineering and geophysical applications is computationally unmanageable because of the need to resolve a wide range of spatiotemporal scales. Large eddy simulation (LES) and Reynolds Averaged Navier-Stokes (RANS) modeling are two most commonly used mathematical modeling frameworks that give accurate predictions by considering the interaction between the unresolved and  grid-resolved scales. The development of these models is termed as the turbulence closure problem and has been a long-standing challenge in fluid mechanics community \cite{durbin1991near, launder1975progress, meneveau2000scale, mellor1982development}.   

In LES, we filter the Navier-Stokes equations using a low-pass filtering operator that separates the motion into small and large scales, and in turn, produces modified equations which are computationally faster to solve than actual Navier-Stokes equations \cite{bardina1980improved, rogallo1984numerical, erlebacher1992toward}. The interaction between grid-resolved and unresolved scales is then taken into account by introducing subgrid-scale stress (SGS) term in the modified equation. The main task of the SGS model is to provide mean dissipation that corresponds to the transfer of energy from resolved scales to unresolved scales (the production of energy at large scales is balanced by the dissipation of energy at small scales based on Kolmogorov's theory of turbulence). The dissipation effect of unresolved scales can be included utilizing an eddy viscosity parameterization obtained through grid-resolved quantities. These approaches are called as functional approaches which assume isotropy of small scales to present the average dissipation of the unresolved scales \cite{frisch1995turbulence}. The most widely used functional model is the Smagorinsky model \cite{smagorinsky1963general} that uses a global constant called the Smagorinsky coefficient to produce mean dissipation of energy. It is observed in many studies that a single value of the Smagorinsky coefficient cannot be used for a variety of flow phenomenon \cite{deardorff1970numerical, mcmillan1980tests, mason1986magnitude, piomelli1988model}. The deficiencies of static Smagorinsky model can be overcome by using dynamic Smagorisnky model (DSM) proposed by Germano et al.\cite{germano1991dynamic}. Lilly \cite{lilly1992proposed} introduced the modification in Germano's DSM by which the stress-strain relationship is optimized with a least-squares approach (we discuss Lilly's version of DSM in detail in Section~\ref{sec:func_models}). Several other versions of Germano's DSM have been proposed, such as localized version to overcome mathematical inconsistencies in standard DSM \cite{ghosal1995dynamic}, Lagrangian version of DSM \cite{meneveau1996lagrangian}, and DSM with a corrector step \cite{park2009reduction}. Even the dynamic procedure is not free from parameter tuning and one has to specify the test filter and grid-filter width ratio to accurately model the SGS stresses. Hence, there is a constant effort to develop a subgrid-scale model that is free from heuristics and can predict the SGS stresses accurately.

In the past decade, the unprecedented amount of data collected from experiments, high-fidelity simulations has facilitated using machine learning (ML) algorithms in fluid mechanics. ML algorithms are now used for flow control, flow optimization, reduce order modeling, flow reconstruction, super-resolution, and flow cleansing \cite{kutz2017deep, brunton2019machine}. One of the first applications of deep learning in fluid mechanics was by Milano \& Koumoutsakos \cite{milano2002neural} who implemented neural network methodology to reconstruct near-wall turbulence and showed an improvement in prediction capability of velocity fields. Subsequently, several ML algorithms such as shallow decoder for flow reconstruction \cite{erichson2019shallow}, a convolutional neural network (CNN) for super-resolution of turbulent flows \cite{fukami2019super}, deep convolutional autoencoder for nonlinear model order reduction \cite{lee2018model} have been proposed. Several studies have been conducted to model the dynamics of chaotic fluid flow using ML algorithms \cite{rudy2017data, long2017pde, raissi2018hidden, pathak2018model, vlachas2018data, raissi2018numerical, pawar2019deep}. Recently there is a growing interest in using the physics of the problem in combination with the data-driven algorithms \cite{raissi2019physics, erichson2019physics, magiera2019constraint, ling2016reynolds, wu2018physics, maulik2019subgrid, pathak2018model, mohebujjaman2019physically}. The physics can be incorporated into these learning algorithms by adding a regularization term (based on governing equations) in loss function or modifying the neural network architecture to enforce certain physical constraints.    

In addition to reduced order modeling and chaotic dynamical systems, the turbulence closure problem has also benefited from the application of ML algorithms and has led to reducing uncertainties in RANS and LES models \cite{duraisamy2019turbulence, lapeyre2019training, king2018deep, wang2018investigations, taira2019revealing}.  Different machine learning algorithms like kernel regression, single hidden layer neural network, random forest \cite{tracey2013application, tracey2015machine, ling2017uncertainty} have been proposed for turbulence closure modeling. Sarghini et al. \cite{sarghini2003neural} proposed the hybrid approach in which the neural network is used for learning Bardina's scale similar subgrid-scale model for turbulent channel flow. Their neural network architecture employed 15 input features consisting of velocity gradients and Reynolds tensor components (made up of fluctuating component of velocity), and turbulent viscosity as the learned variable. The motivation behind this approach was to improve computational performance rather than to learn the true turbulent dynamics. Ling et al. \cite{ling2016reynolds} presented a novel neural network architecture that utilizes a multiplicative layer with an invariant tensor to embed Galilean invariance for the prediction of Reynolds stress anisotropy tensor. Their tensor basis neural network (TBNN) uses five invariants of strain-rate tensor and rotation-rate tensor at a point in the input layer.  In addition to the input layer, the TBNN has tensor input layers that take a tensor basis \cite{pope1975more} (tensor basis includes 10 isotropic basis tensors). They demonstrated the superiority of applying constrained neural network over generic neural network architecture in predicting Reynolds stress anisotropy tensor for various complex flow problems such as duct flow, and wavy channel flow. Maulik et al. \cite{maulik2019subgrid} introduced data-driven turbulence closure framework for subgrid-scale modeling and performed \textit{a priori} and \textit{a posteori} analysis for two-dimensional Kraichnan turbulence. Their neural network architecture employs vorticity, streamfunction, and eddy-viscosity kernel information at nine surrounding grid points to learn the turbulence source term at the central point. They found that the inclusion of eddy viscosity kernels leads to accurate prediction of the turbulence source term. Gamahara  \& Hattori \cite{gamahara2017searching} tested an artificial neural network for finding a new subgrid-scale model in LES of channel flow using the pointwise correlation between grid resolved variables and subgrid stresses. They investigated the effect of different input variables to the neural network and observed that including velocity gradients and vertical distance gives the most accurate prediction for SGS stresses.

Wang et al. \cite{wang2017physics} developed a data-driven framework to learn discrepancies in Reynolds stress models as a function of mean flow features using random forest regression algorithm. They evaluated the performance of the proposed framework in terms of different training and testing parameters for flow characteristics and different geometries. Bhatnagar et al. \cite{bhatnagar2019prediction} built an approximation model using encoder-decoder CNN architecture to determine the aerodynamic flow field around airfoils using the angle of attack, Reynolds number, and airfoil shape as the input variables. Beck et al. \cite{beck2019deep} developed a data-driven approach based on recurrent convolutional neural network for learning the LES closure term for decaying homogeneous isotropic turbulence problem and presented a methodology to construct stable models that can be used in CFD codes. Their architecture includes snapshots of primitive variables and the coarse grid LES operator as input features and unknown subgrid terms in labels. Srinivasan et al. \cite{srinivasan2019predictions} evaluated the capability of multilayer perceptron and long short-term memory network in predicting the turbulent statistics for shear flow.  In the recent work, Pal \cite{pal2019deep} illustrated the two-eight times computational gain that can be attained with a data-driven model that utilizes deep neural network to learn eddy viscosity obtained from the dynamic Smagorinsky model.          

The motivation behind the present work is to address the following questions: which data-driven algorithms are suitable for particular applications, which input features have a significant influence on learning subgrid stresses, which algorithm has better predictive capability, which algorithm is faster, and how much data to use for different ML algorithms for efficient learning? In addition to addressing these questions, we also study the effect of data locality where the information at neighbouring points is found to give improved prediction than point-to-point mapping. The work presented here is concurrent with many of the ideas presented in above studies \cite{maulik2019subgrid, beck2019deep, maulik2017neural, ling2016reynolds, fukami2019super, pal2019deep}, and our main objective is to investigate the performance of different approaches for subgrid-scale modeling in LES of turbulence.  

To achieve these objectives, we examine the performance of data-driven closure models for two-dimensional Kraichnan turbulence \cite{kraichnan1959structure}. Even though the two-dimensional turbulence cannot be realized in practice or experiments but only in numerical simulations, it represents many geophysical flows and provides a starting point in modeling these flows. It finds application in modeling many atmospheric and ocean flows \cite{kraichnan1980two, leith1971atmospheric, boffetta2012two}. A reduction in dimensionality compared to three-dimensional turbulence leads to inverse energy cascade, i.e., the transfer of energy from large scales to smaller scales and direct enstrophy (spatial average of the square of the vorticity) cascade from large scales to small scales \cite{kraichnan1967inertial, batchelor1969computation, leith1971atmospheric}. Therefore, with the presence of complex flow interactions and simplicity of two-dimensional analysis, Kraichnan turbulence will serve as a good testbed for our data-driven closure model analysis. Our approaches are based on three models that employ velocity field, velocity gradient, and the Laplacian of the velocity. These variables are available in any CFD solver and the SGS stresses can be learned in several ways such as point-to-point mapping, neighboring stencil mapping, and learning from the whole field or snapshot. In this work, we demonstrate these different approaches and analyze them in the context of the predictive performance, amount of training data, and computational overhead for training and testing, as well as their data structures.

In Section~\ref{sec:turb_closure}, we introduce the turbulence closure problem and dynamic Smagorinsky model. Section~\ref{sec:neural_network} will present different frameworks investigated in this study. In Section~\ref{sec:numerical_experiments}, we detail the data generation using DNS and will evaluate data-driven turbulence closure models in terms of predictive performance, computational overhead, and data requirement for training.  We demonstrate an additional modeling approach using intelligent eddy viscosity model in Section~\ref{sec:intelligent_nu}
that is computationally faster than the DSM. Finally, we will present the conclusions and future work in Section~\ref{sec:conclusion}. We also describe the hyperparameters selection procedure in Appendix B to obtain optimal neural network architecture.      

\section{Turbulence Closure}
\label{sec:turb_closure}
We begin with the introduction of the turbulence closure framework by outlining governing equations in its primitive variables form used to model incompressible fluid flows. The spatial and temporal evolution of the fluid flow are governed by Navier-Stokes equations which describe the conservation of mass and momentum: 
\begin{align}
    \frac{\partial u_i}{\partial x_i} &= 0, \\
    \frac{\partial u_i}{\partial t} + \frac{\partial u_i u_j}{\partial x_j} &= -\frac{1}{\rho}\frac{\partial p}{\partial x_i} + \nu \frac{\partial }{\partial x_j}\bigg(\frac{\partial u_i}{\partial x_j} + \frac{\partial u_j}{\partial x_i}\bigg),
\end{align}
where $u_i$ is the $i^{\text{th}}$ component of velocity, $p$ is the pressure, and $\nu$ is the kinematic viscosity of fluid. The governing equations for LES (also called as filtered Navier-Stokes equations) are obtained by applying a low-pass filter operation and it results in a grid-filtered system of equations:
\begin{align}
    \frac{\partial \bar{u}_i}{\partial x_i} &= 0, \\
    \frac{\partial \bar{u}_i}{\partial t} + \frac{\partial \overline{u_i u_j}}{\partial x_j} &= -\frac{1}{\rho}\frac{\partial \bar{p}}{\partial x_i} + \nu \frac{\partial }{\partial x_j}\bigg(\frac{\partial \bar{u}_i}{\partial x_j} + \frac{\partial \bar{u}_j}{\partial x_i}\bigg),
\end{align}
where the overbar quantities represent the filtered variables. The filtered Navier-Stokes equations have the nonlinear term $\overline{u_i u_j}$ which is unknown due to truncation of small eddies by spatial filtering operation. The decomposition of nonlinear term \cite{leonard1975energy} can be given as 
\begin{equation}\label{eq:sgs}
    \overline{u_i u_j} = \tau_{ij} + \bar{u}_i \bar{u}_j,
\end{equation}
where $\tau_{ij} = \overline{u_i u_j} -  \bar{u}_i \bar{u}_j$ is the subgrid-scale stress that consists of cross-stress tensor (which represents interaction between large and small scales), Reynolds subgrid tensor (which represents interaction between subgrid scales), and Leonard tensor (which represents the interactions among large scales). Using this decomposition the filtered Navier-Stokes equation can be written as
\begin{equation}
    \frac{\partial \bar{u}_i}{\partial t} + \frac{\partial \bar{u}_i\bar{u}_j}{\partial x_j} = -\frac{1}{\rho}\frac{\partial \bar{p}}{\partial x_i} + \nu \frac{\partial }{\partial x_j}\bigg(\frac{\partial \bar{u}_i}{\partial x_j} + \frac{\partial \bar{u}_j}{\partial x_i}\bigg) - \frac{\partial \tau_{ij}}{\partial x_j}.
\end{equation}
The main challenge in subgrid-scale modeling is to approximate this $\tau_{ij}$ term and the approximated model should provide sufficient dissipation corresponding to transfer of energy from large eddies to unresolved eddies. The static Smagorinsky model \cite{smagorinsky1963general} which uses an effective eddy viscosity to model SGS stresses is given by 
\begin{equation}
    \tau_{ij}^{M,d} = -2(C_s \Delta)^2 |\bar{S}|\bar{S}_{ij},
    \label{eq:stat_smag}
\end{equation}
where the superscript $M$ stands for the model, $d$ means the deviatoric (traceless) part of the tensor, $\Delta$ is the grid-filter width, and $C_s$ is the static Smagorinsky coefficient. A derivation of Smagorinsky model for two-dimensional case is provided in Appendix A. The terms $|\bar{S}|$ and $\bar{S}_{ij}$ in the above equation are calculated as
\begin{equation}
    \bar{S}_{ij} = \frac{\partial \bar{u}_i}{\partial x_j} + \frac{\partial \bar{u}_j}{\partial x_i}, \quad |\bar{S}| = \sqrt{2 \bar{S}_{ij} \bar{S}_{ij}}.
    \label{eq:S}
\end{equation}
It should be noted that the static Smagorinsky model might be excessive or under dissipative with suboptimal values of $C_s$. 
It was found in many studies that the Smagorinsky coefficient is different for different flows and additional modifications are needed in the near-wall region \cite{deardorff1970numerical, mcmillan1980tests, mason1986magnitude, piomelli1988model}. To tackle this problem, the dynamic Smagorinksy model \cite{germano1991dynamic, lilly1992proposed} was introduced that allowed the $C_s$ to be computed dynamically based on the flow, time, resolution, and spatial location. The dynamic Smagorinsky model is discussed in Section~\ref{sec:func_models}. 

\subsection{Dynamic Smagorinsky Model}
\label{sec:func_models}
Germano et al. \cite{germano1991dynamic} introduced the dynamic procedure that calculates the Smagorinsky coefficient based on the local flow structure dynamically instead of assuming a constant value. The dynamic procedure consists of applying a secondary spatial filter called as the test filter to the grid-filtered Navier-Stokes equations. The test filtered equations can be written as  
\begin{equation}
    \frac{\partial \hat{\bar{u}}_i}{\partial t} + \frac{\partial \hat{\bar{u}}_i \hat{\bar{u}}_j}{\partial x_j} = -\frac{1}{\rho}\frac{\partial \hat{\bar{p}}}{\partial x_i} + \nu \frac{\partial }{\partial x_j}\bigg(\frac{\partial \hat{\bar{u}}_i}{\partial x_j} + \frac{\partial \hat{\bar{u}}_j}{\partial x_i}\bigg) - \frac{\partial \mathcal{T}_{ij}}{\partial x_j},
\end{equation}
 where the caret over the overbar represents the test filtered variables. The test filtered subgrid stress $\mathcal{T}_{ij}$ (also called as subtest-scale stress) is given by
 \begin{equation}
     \mathcal{T}_{ij} = \widehat{\overline{u_i u_j}} - \hat{\bar{u}}_i \hat{\bar{u}}_j. 
 \end{equation}
Similar to Equation~\ref{eq:stat_smag}, the subtest-scale stress can be approximated as 
\begin{equation}
    \mathcal{T}_{ij}^{M,d} = -2(C_s \hat{\Delta})^2 |\hat{\bar{S}}|\hat{\bar{S}}_{ij},
\end{equation}
where $\hat{\Delta}$ is the test filter scale. The application of dynamic procedure leads to introduction of grid filtered SGS stress given by
\begin{align}
    \mathcal{L}_{ij} &= \mathcal{T}_{ij} - \hat{\tau}_{ij}, \\
    &= \widehat{\bar{u}_i \bar{u}_j} - \hat{\bar{u}}_i \hat{\bar{u}}_j.
\end{align}
In dynamic procedure, the value of $C_s$ is chosen in such a way that the error (also called as Germano identity error) given in the below equation is minimized 
\begin{align}\label{eq:gemano_error}
    \epsilon_{ij} &= \mathcal{T}^{M,d}_{ij} - \hat{\tau}_{ij}^{M,d} - \mathcal{L}_{ij}^{d}, \\ \label{eq:test_cs}
    &= -2(C_s\hat{\Delta})^2 |\hat{\bar{S}}|\hat{\bar{S}}_{ij} + 2[\reallywidehat{(C_s \Delta)^2|\bar{S}|\bar{S}_{ij}}] - \mathcal{L}_{ij}^{d}.
\end{align}
The computation of $C_s$ in the above equation that minimizes the Germano identity error is not straight-forward as Equation~\ref{eq:gemano_error} is a tensor equation (three equations for two-dimensional flows) for only one unknown $C_s$. Also, the coefficient $C_s$ in the second term of Equation~\ref{eq:test_cs} is inside the test filter operator. However, it is often approximated as
\begin{align}\label{eq:gemano_error2}
    \epsilon_{ij} = -2(C_s\hat{\Delta})^2 |\hat{\bar{S}}|\hat{\bar{S}}_{ij} + 2(C_s \Delta)^2\reallywidehat{|\bar{S}|\bar{S}_{ij}} - \mathcal{L}_{ij}^{d},
\end{align}
which makes the formulation mathematically consistent only when $C_s$ is a constant valued variable.  
\begin{equation}
    (C_s \Delta)^2 = \frac{\mathcal{M}_{ij} \mathcal{L}_{ij}^{d}}{\mathcal{M}_{ij} \mathcal{M}_{ij}},\label{eq:lillys_model}
\end{equation}
where
\begin{equation}
    \mathcal{M}_{ij} = 2\reallywidehat{|\bar{S}|\bar{S}_{ij}} - 2\bigg(\frac{\hat{\Delta}}{\Delta}\bigg) |\hat{\bar{S}}|\hat{\bar{S}}_{ij}. 
\end{equation}
From the original dynamic Smagorinsky model \cite{germano1991dynamic}, it was found that the denominator in Equation~\ref{eq:lillys_model} can become very small leading to excessively large value of $C_s$. Furthermore, Equation~\ref{eq:lillys_model} becomes mathematically ill-posed since we factor $C_s$ from the convolution filter (i.e., see Equation~\ref{eq:gemano_error2}). Therefore some type of averaging is necessary in practice as given below 
\begin{equation}
    (C_s \Delta)^2 = \frac{<\mathcal{M}_{ij} \mathcal{L}_{ij}^{d}>_{h}^{+}}{<\mathcal{M}_{ij} \mathcal{M}_{ij}>_{h}},
    \label{eq:cs2}
\end{equation}
where $<\cdot>_h$ denotes the spatial averaging, and $<\cdot>_h^{+} = 0.5(<\cdot> + |<\cdot>|)$ denotes the positive clipping. The above averaging gives a global value of $C_s$ which changes over time. One of the advantage of dynamic Smagorinsky model is that the numerator can also take negative values corresponding to backscatter, i.e., transfer of energy from small scales to large scales. If the averaging is not done, the dynamic model leads to a highly variable eddy viscosity field and can cause numerical simulations to become unstable \cite{liu1995experimental, meneveau2000scale}. These findings are also applicable to data-driven turbulence closure modeling as demonstrated in recent studies \cite{maulik2019subgrid, beck2019deep}. 

\section{Data-driven Turbulence Closure}
\label{sec:neural_network}
In this section, we outline different data-driven turbulence closure frameworks investigated in this work. As discussed in Section~\ref{sec:turb_closure}, we try to approximate $\tau_{ij}$ using resolved flow variables on coarse grid in subgrid-scale modeling. We can consider this as a regression problem that can be studied using various classes of supervised machine learning algorithms. In the case of supervised algorithms, we try to learn the optimal map between inputs and outputs. We focus on two algorithms: an artificial neural network (ANN) also called as multilayer perceptron and convolutional neural network (CNN) to build data-driven closure models. 

An artificial neural network consists of several layers made up of the predefined number of nodes (also called as neurons). A node combines the input from the data with a set of coefficients called weights. These weights either amplify or dampen the input and thereby assign the significance to the input in relation to the output that the ANN is trying to learn. In addition to the weights, these nodes have a bias for each input to the node. The input-weight product and the bias are summed and this sum is passed through a node's activation function. The activation function introduces nonlinearity and this allows the neural network to map complex relations between the input and output. The above process can be described using the matrix operation as given by \cite{hagan1996neural}
\begin{equation}
    S^l = \mathbf{W}^lX^{l-1},
\end{equation}
where $X^{l-1}$ is the output of the $(l-1)$th layer, $\mathbf{W}^l$ is the matrix of weights for the $l$th layer. The output of the $l$th layer is given by
\begin{equation}
    X^l = \zeta(S^l+B^l),
\end{equation}
where $B^l$ is the vector of biasing parameters for the $l$th layer and $\zeta$ is the activation function. If there are $L$ layers between the input and the output, then the mapping of the input to the output can be derived as follow
\begin{equation}
    \tilde{Y} =\zeta_L(\mathbf{W}^L,B^L,\dots,\zeta_2(\mathbf{W}^2,B^2,\zeta_1(\mathbf{W}^1,B^1,X))),
\end{equation}
where $X$ and $\tilde{Y}$ are the input and output of the ANN, respectively. 

The matrix $\mathbf{W}$ and $B$ are optimized through backpropagation and some optimization algorithm. The backpropagation algorithm provides a way to compute the gradient of the objective function efficiently and the optimization algorithm gives a rapid way to learn optimal weights. For the regression problem, usually, the objective is to learn the weights associated with each node in such a way that the root mean square error between the true labels $Y$ and output of the neural network $\tilde{Y}$ is minimized. The backpropagation algorithm proceeds as follows: (i) the input and output of the neural network are specified along with some initial weights, (ii) the training data is run through the network to produce output $\tilde{Y}$ whose true label is $Y$, (iii) the derivative of the objective function with each of the training weight is computed using the chain rule, (iv)  the weights are updated based on the learning rate and then we go to step (ii). We continue to iterate through this procedure until convergence or the maximum number of iterations is reached. There are a number of ways in which the weights can be initialized \cite{glorot2010understanding}, the optimization algorithm is selected \cite{sutskever2013importance,kingma2014adam,ruder2016overview}, and the loss function be regularized \cite{wan2013regularization, srivastava2014dropout} either to speed up the learning process or to prevent overfitting. Furthermore, highly nonlinear relationship between the input and output (as in the case of turbulence) necessitates the need of deep neural network architecture which are prone to overfitting. Pruning neural network weights can significantly reduce the parameter count leading to better generalization \cite{bartoldson2019generalization}.    

\begin{figure*}[htbp]
\centering
{\includegraphics[width=0.8\textwidth]{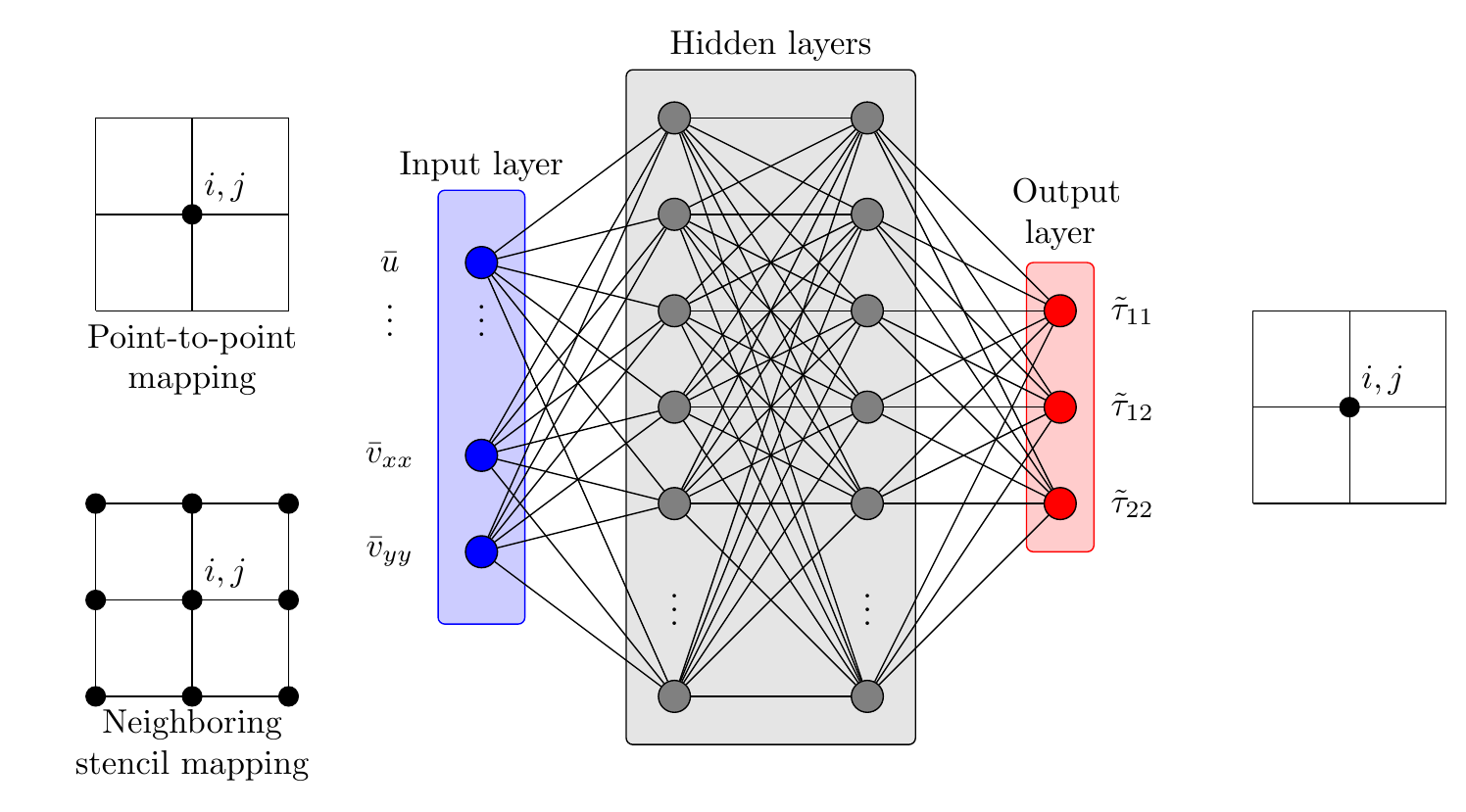}}
\caption{Feedforward neural network for point-to-point and neighboring stencil mapping of resolved variables to SGS stresses.}
\label{fig:dnn}
\end{figure*}

It has been demonstrated in many studies how an ANN can be used for learning input-output relationship in the context of turbulence closure modeling \cite{ling2016reynolds, maulik2019subgrid, maulik2017neural, wang2017physics, beck2019deep, srinivasan2019predictions, zhu2019machine, xie2019artificial, yang2019predictive}. We use two types of mapping using ANN as shown in Figure~\ref{fig:dnn}. The first one is point-to-point mapping in which only the information at a point is used to learn the SGS stresses at that point. We can include different features at that point and evaluate its predictive capability by means of probability density function based analysis. We use three classes of point-to-point mapping in our data-driven closure models which are given below
\begin{align} 
    \mathbb{M}1 : \{\bar{u}, \bar{v}\} \in \mathbb{R}^2 &\rightarrow \{ \tilde{\tau}_{11}, \tilde{\tau}_{12}, \tilde{\tau}_{22}\} \in \mathbb{R}^3, \label{eq:m1} \\
    \mathbb{M}2 : \{\bar{u}, \bar{v}, \bar{u}_x, \bar{u}_y, \bar{v}_x, \bar{v}_y\} \in \mathbb{R}^6 &\rightarrow \{ \tilde{\tau}_{11}, \tilde{\tau}_{12}, \tilde{\tau}_{22}\} \in \mathbb{R}^3, \label{eq:m2} \\
    \mathbb{M}3 : \{\bar{u}, \bar{v}, \bar{u}_x, \bar{u}_y, \bar{v}_x, \bar{v}_y, \bar{u}_{xx}, \bar{u}_{yy}, \bar{v}_{xx}, \bar{v}_{yy}\} \in \mathbb{R}^{10} &\rightarrow \{ \tilde{\tau}_{11}, \tilde{\tau}_{12}, \tilde{\tau}_{22}\} \in \mathbb{R}^3, \label{eq:m3}
\end{align}
where $\bar{u}$, $\bar{v}$ are the the velocities in $x$ and $y$ direction, the subscript $x$ and $y$ denote the first-derivative, the subscript $xx$ and $yy$ are the second-derivative, and $\tilde{\tau}_{11}, \tilde{\tau}_{12}, \tilde{\tau}_{22}$ are the approximated SGS stresses.  

The second approach is to use the information at neighbouring points to learn SGS stresses at a point. We can either use information of just north, south, east, and west points or information at all nine neighbouring points. In our neighboring stencil mapping, we use information at nine grid points. As we will see in Section~\ref{sec:numerical_experiments}, one of the advantages of this approach is that the ANN can learn the input-output mapping with less number of input features. Similar to point-to-point mapping, we use three classes of input features for neighboring stencil mapping. Therefore, in case of neighboring stencil mapping, we will have nine times the number of input features as in case of point-to-point mapping.  

In addition to ANN, we also investigate CNN for subgrid-scale modeling. CNNs have been found to perform better than ANNs when the data is in the form of snapshots such as images and is widely used for computer vision tasks such as object detection \cite{krizhevsky2012imagenet,ren2015faster}, and improving the quality of images \cite{kim2016accurate, dong2016accelerating}. CNNs have also been successfully applied for detecting flow disturbances \cite{hou2019machine}, super-resolution analysis of turbulent flow field \cite{fukami2019super}, and turbulence closure modeling \cite{beck2019deep, lapeyre2019training, nikolaou2018modelling, nikolaou2019progress}. One of the differences between ANN and CNN is that the training sample to the CNN is not given as one-dimensional vector but as a two-dimensional snapshot image. This will preserve the original multi-dimensional structure and will aid in learning the SGS stresses. Apart from that, the number of parameters to be learned in CNN is significantly less than ANN due to parameter sharing scheme.

The Conv layers are the fundamental building blocks of CNN. Similar to weights in case of ANN, Conv layers have filters, also called as kernels that has to be learned using the backpropagation algorithm. The filter has a smaller shape but it extends in through the full depth of the input volume of previous layer. For example, if the input to the CNN has $64 \times 64 \times 3$ dimension where 3 is the number of input features, the kernels of first Conv layer can have $3 \times 3 \times 3$ shape. During the forward propagation, we convolve the filter across the width and height of the input volume to produce the two-dimensional map. The two-dimensional map is constructed by computing the dot product between the entries of the filter and the input volume at any position and then sliding it over the whole volume. Mathematically the convolution operation corresponding to one filter can be given as 
\begin{equation}
    S^l_{ij} = \sum_{p=-\Delta_i/2}^{\Delta_i/2} \sum_{q=-\Delta_j/2}^{\Delta_j/2} \sum_{r=-\Delta_k/2}^{\Delta_k/2} \mathbf{W}_{pqr}^l X_{i+p ~ j+q ~ k+r}^{l-1} + B_{pqr},
\end{equation}
where $\Delta_i$, $\Delta_j$, $\Delta_k$ are the sizes of filter in each direction, $\mathbf{W}_{pqr}^l$ are the entries of the filter for $l$th Conv layer, $ B_{pqr}$ is the biasing parameter, and $X_{ijk}^{l-1}$ is the input from $(l-1)$th layer. Each Conv layer will have a set of predefined filters and the two-dimensional map produced by each filter is then stacked in the depth dimension to produce a three-dimensional output volume. This output volume is passed through an activation function to produce a non-linear map between inputs and outputs. The output of the $l$th layer is givn by 

\begin{equation}
    X_{ijk}^l = \zeta(S_{ijk}^l),
\end{equation}
where $\zeta$ is the activation function. It should be noted that as we convolve the filter across the input volume, the size of the input volume shrinks in height and width dimension. Therefore, it is common practice to pad the input volume with zeros called as zero-padding. The zero-padding allows us to control the shape of the output volume and is used in our data-driven closure framework to preserve the shape so that input and output width and height are same. The size of the zero-padding is an additional hyperparameter in CNN.  

\begin{figure*}[htbp]
\centering
{\includegraphics[width=0.8\textwidth]{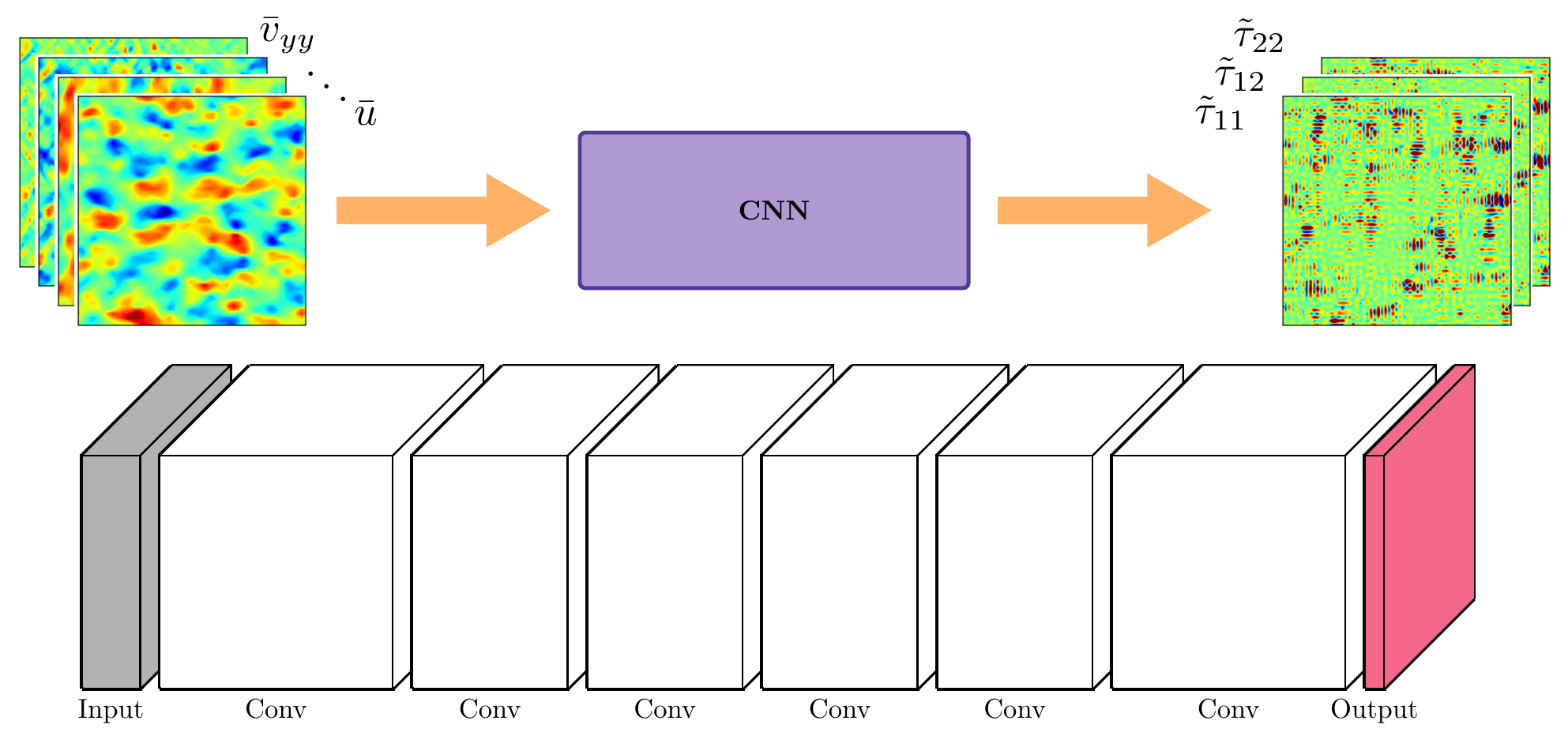}}
\caption{Convolution neural network for mapping of resolved variables to SGS stresses. Our CNN architecture is fairly simple and we use zero padding to keep the shape same as we go from input to the output. }
\label{fig:cnn}
\end{figure*}

Figure~\ref{fig:cnn} shows the schematic of the CNN architecture used in our data-driven closure framework. The input to the CNN is obtained by stacking snapshots of resolved variables and their derivatives at the coarse grid. Similar to ANN, we use three classes of input features as given in Equations~\ref{eq:m1} -~\ref{eq:m3}. Therefore, for model $\mathbb{M}1$, each sample of the input volume will have $64 \times 64 \times 2$ shape and the the sample of output volume will have $64 \times 64 \times 3$ shape. 

\section{Intelligent SGS Modeling}
\label{sec:numerical_experiments}
The present study is focused on the comparison of data-driven closure approaches discussed in Section~\ref{sec:neural_network} for SGS modeling. We use two-dimensional Kraichnan turbulence problem as our prototype example to show the comparison of different frameworks. The purpose of this test problem is to see how the abundant population of randomly generated vortices evolve \cite{tabeling2002two}. For data-driven frameworks, we use true subgrid-scale stresses ($\tau_{ij}$) generated by solving the two-dimensional Navier-Stokes equation with DNS. The computational domain is square in shape with the dimension $[0,2\pi] \times [0,2 \pi]$ in $x$ and $y$ directions. The domain has the periodic boundary condition in $x$ and $y$ directions. We use pseudo-spectral solver for DNS of Kraichnan turbulence problem. The pseudo-spectral solver is accurate in a sense that it does not introduce any discretization error. We use hybrid explicit third-order Runge-Kutta scheme and implicit Crank-Nicolson scheme for the time integration. It should be noted that we solve the Navier-Stokes equations using streamfunction-vorticity formulation and then compute primitive variables using a spectral method for differentiation. The streamfunction-vorticity formulation eliminates the pressure term from the momentum equation and hence, there is no odd-even coupling between the pressure and velocity. This allows us to use collocated grid instead of the staggered grid.     

\begin{figure*}[htbp]
\centering
{\includegraphics[width=0.95\textwidth]{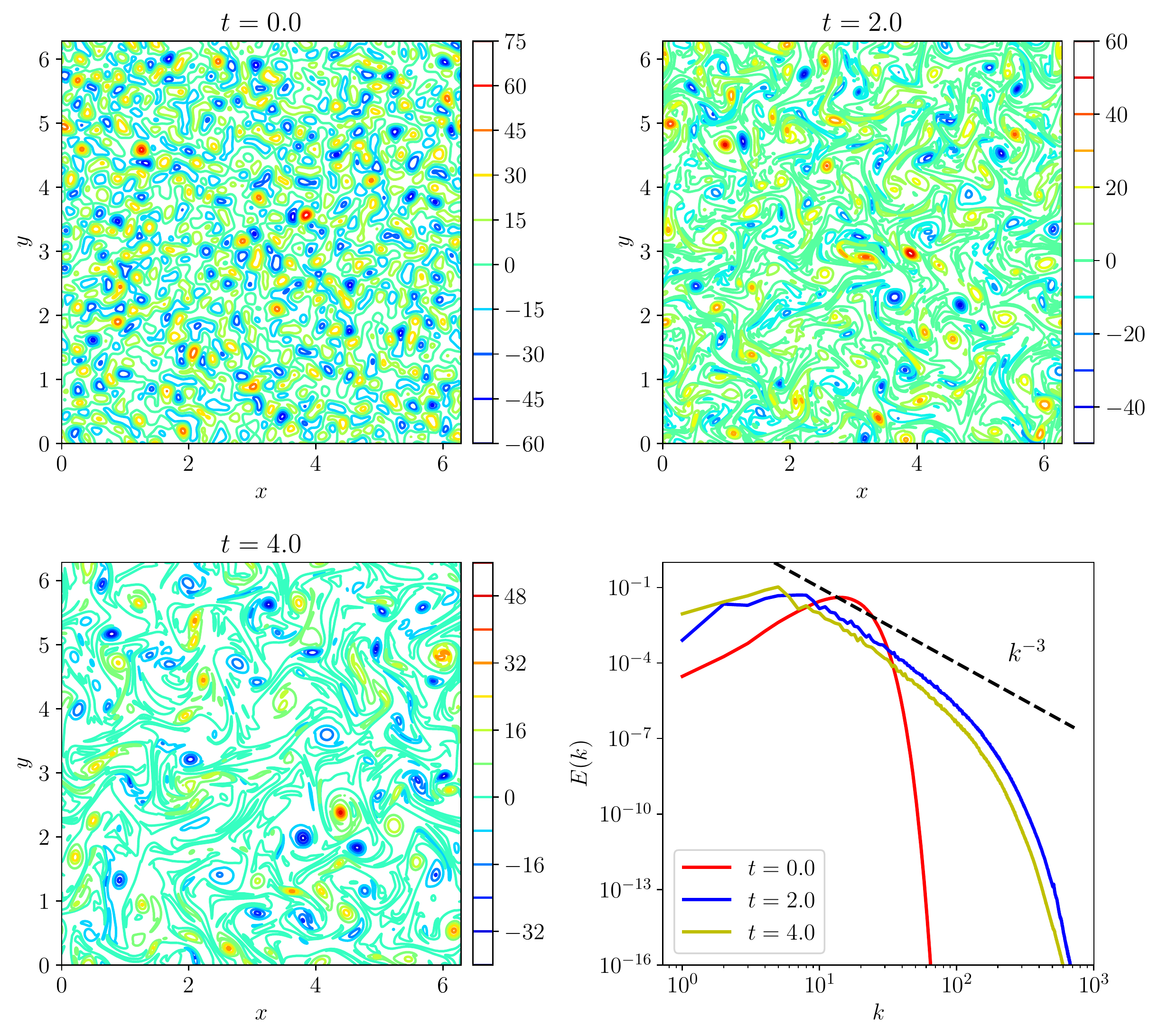}}
\caption{Time evolution of the vorticity field and energy spectrum from time $t=0.0$ to $t=4.0$ for Re=4000 at grid resolution $1024 \times 1024$.}
\label{fig:field_spectral}
\end{figure*}

The DNS solution is computed for Re = 4000 with the grid resolution of $1024 \times 1024$. We integrate the solution from time $t=0$ to $t=4$ with $\Delta t  = 1\times 10^{-3}$. The evolution of the vorticity field and the energy spectrum for two-dimensional Kraichnan turbulence are shown in Figure~\ref{fig:field_spectral}. The initial condition for the energy spectrum is assigned in such a way that the maximum value of energy is designed to occur at the wavenumber $k=10$. Using this energy spectrum and random phase function, the initial vorticity field is assigned. The random vorticity field assigned is kept identical (using constant seed) in all our numerical experiments for comparison and reproducing the results. Interested readers are referred to related work \cite{orlandi2012fluid, san2012high} for the energy spectrum equation and randomization process. We collect 400 snapshots of data from time $t=0$ to $t=4$. The Kraichnan-Batchelor-Leith (KBL) theory states that the energy spectrum of two-dimensional turbulence is proportional to $k^{-3}$ in the inertial range and we observe this behavior with our numerical solution at $t=2.0$ and $t=4.0$ as shown in Figure~\ref{fig:field_spectral}. For LES, we coarsen the solution on $64 \times 64$ grid resolution using the cut-off filter. The resolved flow variables at the coarse grid are then used to compute input features for data-driven turbulence closure models. 

\begin{figure*}[htbp]
\centering
{\includegraphics[width=0.8\textwidth]{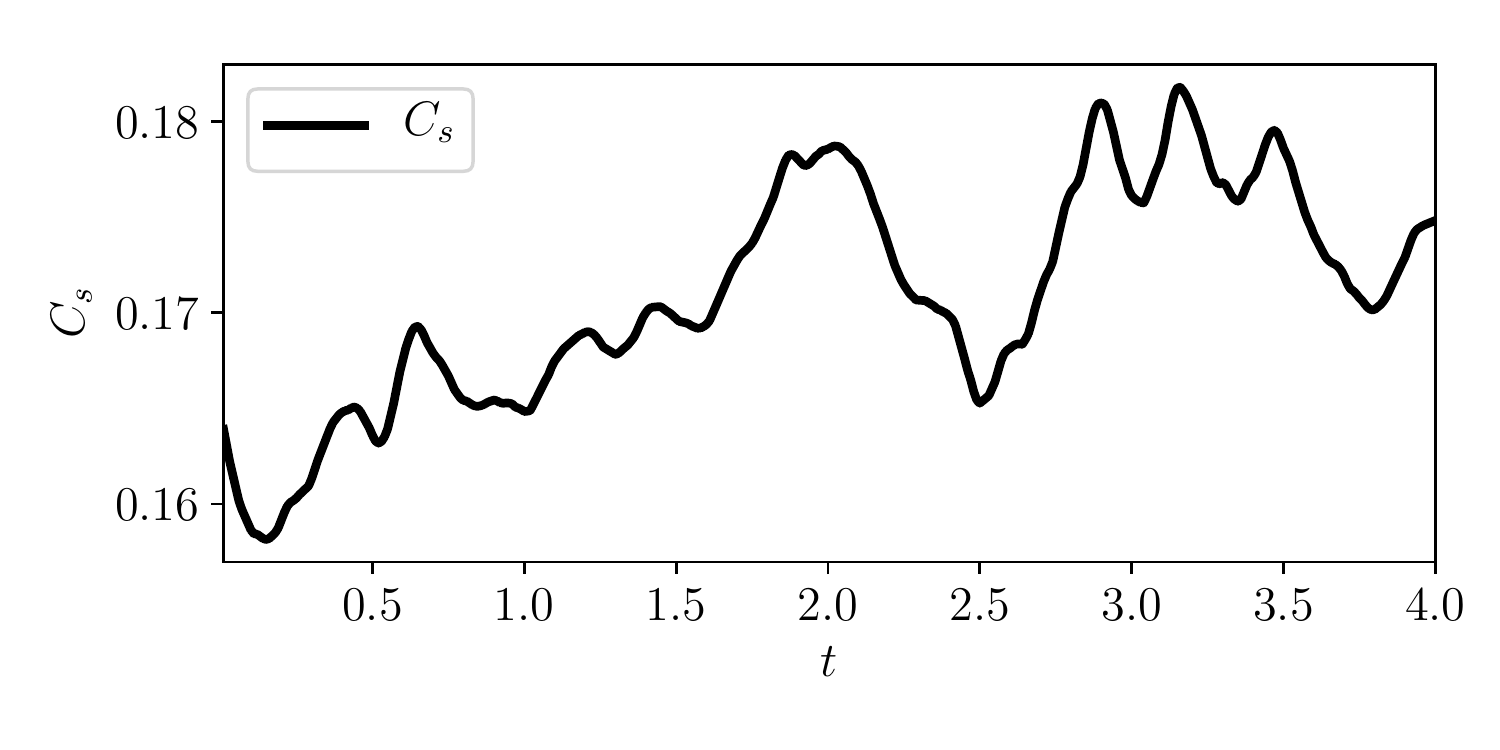}}
\caption{Evolution of the Smagorinsky coefficient ($C_s$) for two-dimensional Kraichnan turbulence problem computed using Lily's version of dynamic model with positive clipping.}
\label{fig:cs2}
\end{figure*}

We analyze the performance of data-driven closure models against the dynamic Smagorinsky model discussed in Section~\ref{sec:func_models}. One of the advantages of DSM is that the Smagorinsky coefficient is computed using the resolved field variables in a dynamic fashion and does not require \textit{a priori} coefficient specification. Due to this advantage, DSM is widely used in LES of engineering and geophysical applications \cite{kleissl2006numerical, galperin1993large, khani2015large, moin1991dynamic}. The only parameter that has to be specified for the DSM is the filter width ratio (i.e., a ratio between the test and grid filters). We use the low-pass spatial filter as a test filter and the test filter scale is $\hat{\Delta}=2\Delta$. Figure~\ref{fig:cs2} shows the temporal evolution of Smagorinsky coefficient from time $t=0.0$ to $t=4.0$ computed with DSM. Ths Smagorinsky coefficient changes between 0.16 to 0.18. We have to use this low-pass filtering operation eight times for the DSM and the procedure becomes computationally expensive compared to static Smagorinsky model. A data-driven turbulence closure model can also be developed to learn dynamic eddy viscosity (computed by DSM) instead of learning true SGS stresses. The similar approach was implemented by Sarghini et al. \cite{sarghini2003neural} for learning Bardina's scale similar subgrid-scale model to improve computational performance. We use the similar framework for learning eddy viscosity computed by the DSM, and is detailed in Section~\ref{sec:intelligent_nu}. The DNS code for the pseudo-spectral solver, and the code for DSM is implemented using vectorized Python. This will allow us to compare the computational performance of the DSM with data-driven closure models fairly (the most popular libraries for machine learning like Keras, Tensorflow are available in Python). 

We use two metrics to determine the performance of data-driven closure models. First one is the cross-correlation between true SGS stresses and the predicted SGS stresses. The cross-correlation ($cc$) is calculated using below formula 

\begin{equation}
    cc = \frac{\text{cov}(Y,\tilde{Y})}{\sigma_{Y} \sigma_{\tilde{Y}}},
\end{equation}
where the covariance (cov) is defined as 
\begin{equation}
    {\text{cov}}(Y,\tilde{Y}) = E[(Y-E[Y])(\tilde{Y}-E[\tilde{Y}])].
\end{equation}
In above equations, $Y$ is the true field, $\tilde{Y}$ is the predicted field, $\sigma_Y$ is the standard deviation of $Y$, $\sigma_{\tilde{Y}}$ is the standard deviation of ${\tilde{Y}}$, $E[Y]$ is the expected value of the true field, and $E[\tilde{Y}]$ is the expected value of the predicted field. The expected value and the standard deviation for a sample field $Y$ can be given as 
\begin{equation}
    E[Y] = \frac{\sum_{i=1}^n y_i}{n}, \quad \sigma_Y=\sqrt{\frac{\sum_{i=1}^n (y_i - E[Y])^2}{n}}.
\end{equation}

In addition to the cross-correlation, we assess the model's performance using probability density function (PDF) based analysis. We test all data-driven closure models using 350 snapshots of training data from time $t=0.0$ to $t=3.5$. We use 20\% of the training data for validation. We use the resolved field variables at time $t=4.0$ to determine SGS stresses as out-of-training data snapshot. This data has not been seen by the neural network during training and hence the model's performance should be measured against this data. In addition to using 350 data snapshots for training, we test extremely sub-sampled data with 70, and 24 snapshots. This will help us in understanding how much data to use for each of these data-driven closure models for learning true SGS stresses efficiently. The hyperparameters for all neural network architectures are selected using gridsearch algorithm coupled with five-fold cross-validation and the procedure is discussed in detail in Appendix B. We have to be cautious when measuring CPU time for deployment of trained model and the sample code for CPU time measurement is given in Appendix C. The sample code for our ANN and CNN architecture is provided in Appendix D.  

\subsection{Point-to-point Mapping} \label{sec:direct_injection}
\begin{figure*}[htbp]
{\includegraphics[width=0.98\textwidth]{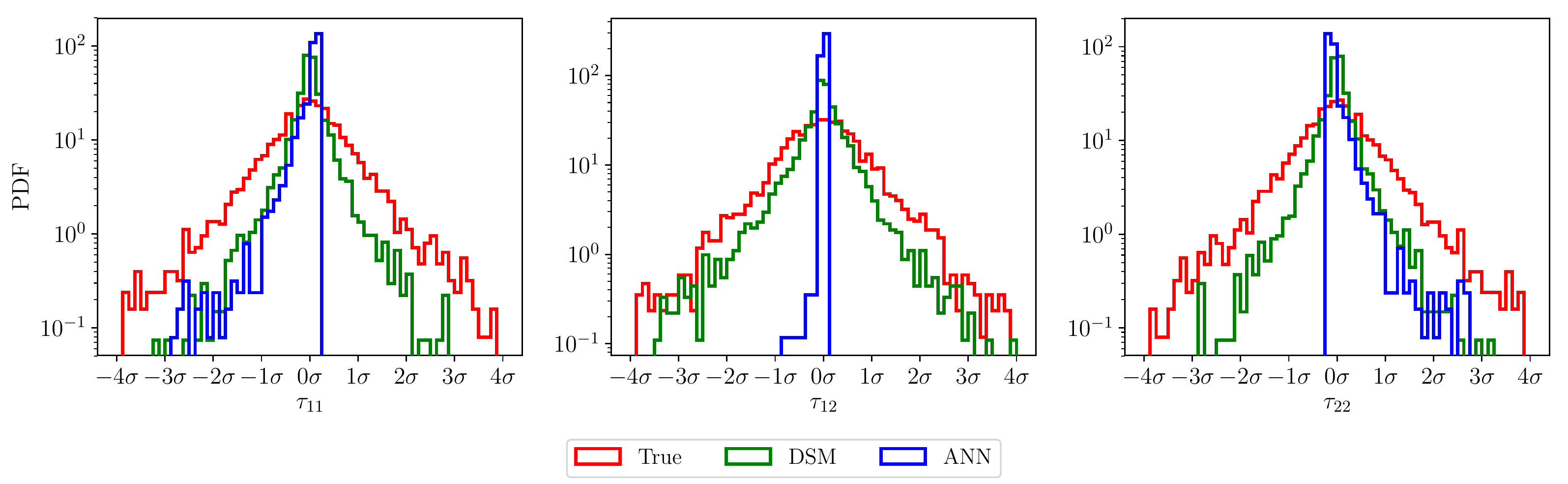}}
\caption{Probability density function for SGS stress distribution with point-to-point mapping. The ANN is trained using $\mathbb{M}1 :\{{\bar{u},\bar{v}}\} \rightarrow \{\tilde{\tau}_{11},\tilde{\tau}_{12},\tilde{\tau}_{22}\}$. The training set consists of 350 time snapshots from time $t=0.0$ to $t=3.5$ and the model is tested for $400{th}$ snapshot at $t=4.0$.}
\label{fig:ts_dnn_di_1_2_350}
\end{figure*}

\begin{figure*}[htbp]
{\includegraphics[width=0.98\textwidth]{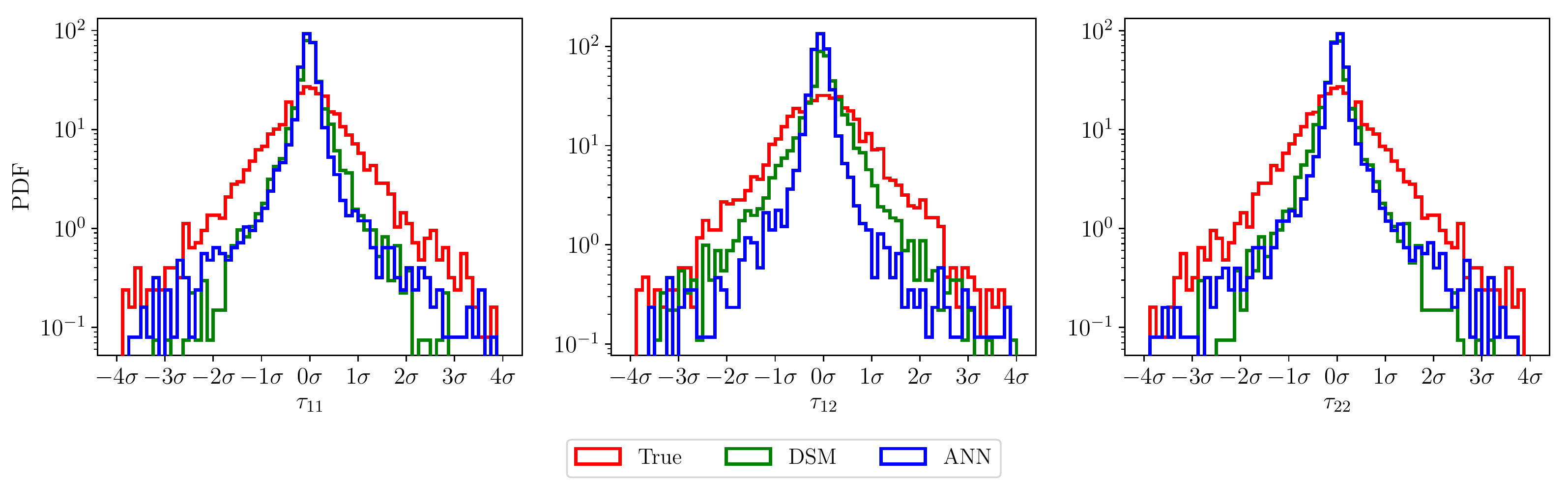}}
\caption{Probability density function for SGS stress distribution with point-to-point mapping. The ANN is trained using $\mathbb{M}2 :\{{\bar{u},\bar{v},\bar{u}_x,\bar{u}_x,\bar{u}_y,\bar{v}_y}\} \rightarrow \{\tilde{\tau}_{11},\tilde{\tau}_{12},\tilde{\tau}_{22}\}$. The training set consists of 350 time snapshots from time $t=0.0$ to $t=3.5$ and the model is tested for $400{th}$ snapshot at $t=4.0$.}
\label{fig:ts_dnn_di_1_6_350}
\end{figure*}

\begin{figure*}[htbp]
{\includegraphics[width=0.98\textwidth]{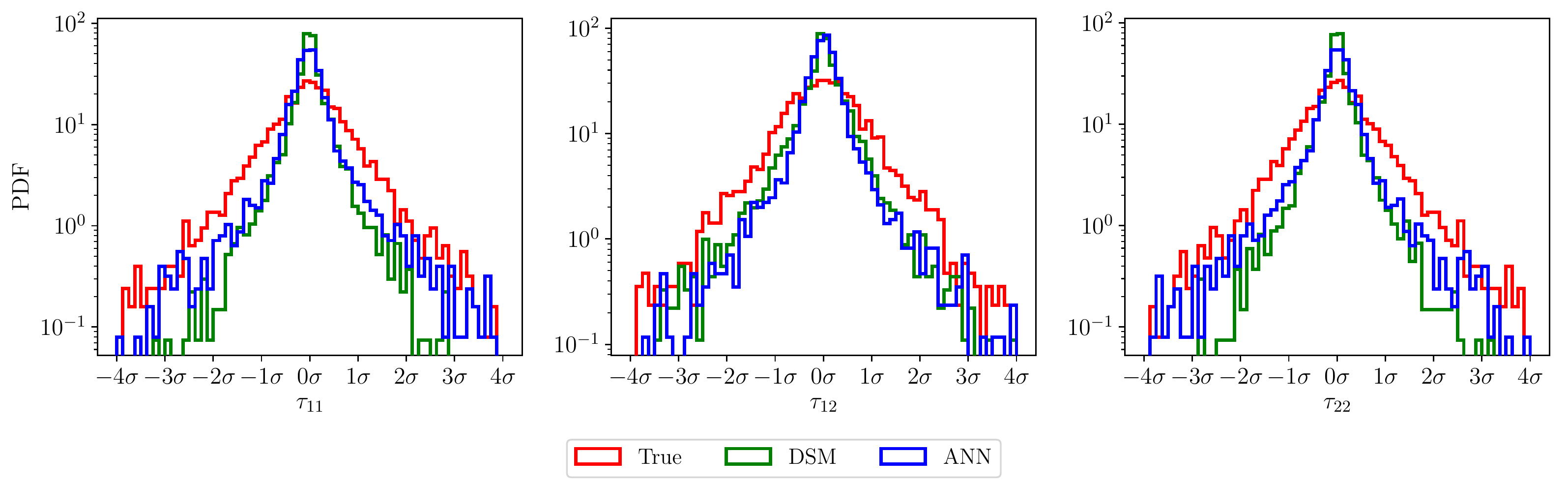}}
\caption{Probability density function for SGS stress distribution with point-to-point mapping. The ANN is trained using $\mathbb{M}3 :\{{\bar{u},\bar{v},\bar{u}_x,\bar{u}_x,\bar{u}_y,\bar{v}_y}, \bar{u}_{xx}, \bar{u}_{yy}, \bar{v}_{xx}, \bar{v}_{yy}\} \rightarrow \{\tilde{\tau}_{11},\tilde{\tau}_{12},\tilde{\tau}_{22}\}$. The training set consists of 350 time snapshots from time $t=0.0$ to $t=3.5$ and the model is tested for $400{th}$ snapshot at $t=4.0$.}
\label{fig:ts_dnn_di_1_10_350}
\end{figure*}

\begin{figure*}[htbp]
{\includegraphics[width=0.98\textwidth]{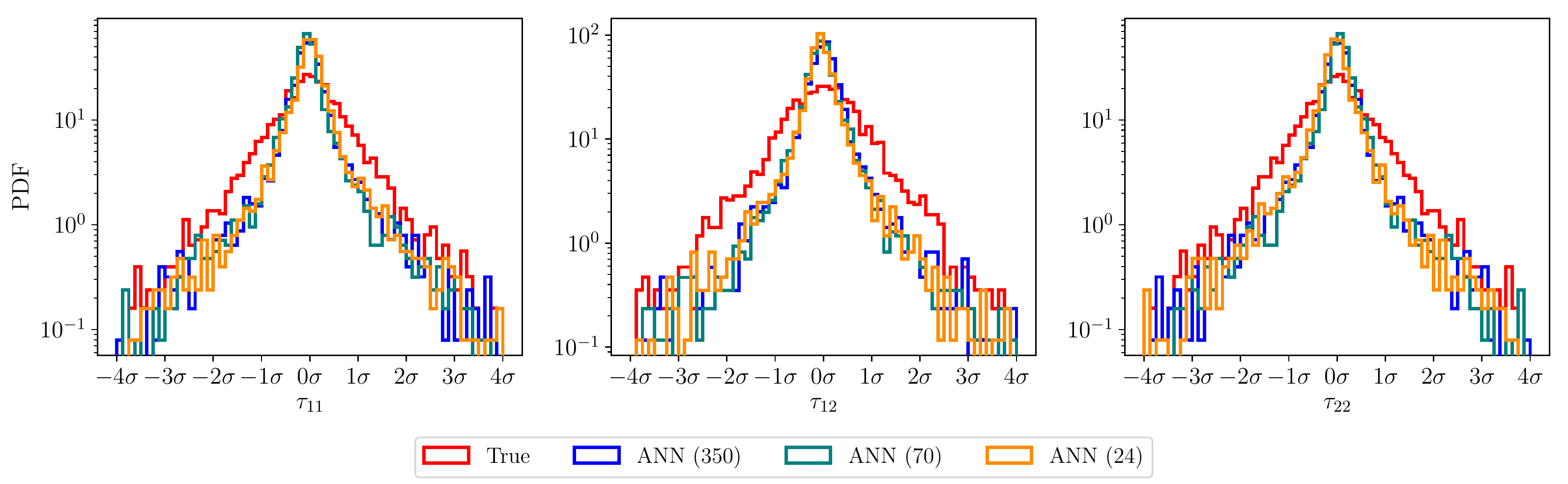}}
\caption{Probability density function for SGS stress distribution with point-to-point mapping. The ANN is trained using $\mathbb{M}3 :\{{\bar{u},\bar{v},\bar{u}_x,\bar{u}_x,\bar{u}_y,\bar{v}_y}, \bar{u}_{xx}, \bar{u}_{yy}, \bar{v}_{xx}, \bar{v}_{yy}\} \rightarrow \{\tilde{\tau}_{11},\tilde{\tau}_{12},\tilde{\tau}_{22}\}$. The model is trained using different number of snapshots between $t=0.0$ to $t=3.5$ and the model is tested for $400{th}$ snapshot at $t=4.0$. }
\label{fig:ann_p2p_ns}
\end{figure*}

We first discuss the performance of point-to-point mapping ANN in predicting true SGS stresses. Table~\ref{tab:p2p_ann} gives the cross-correlation between true and predicted SGS stresses for three different models (i.e., models $\mathbb{M}1$, $\mathbb{M}2$, $\mathbb{M}3$ presented in Section~\ref{sec:neural_network}). The cross-correlation between the DSM and true stresses is low because the DSM model cannot capture the phase correctly (as we will see at the end of this section). It can be seen that the correlation between true and predicted SGS stresses is very poor when we use only coarse-grid resolved velocities at a point to determine the stresses at that point. It is clear from Figure~\ref{fig:ts_dnn_di_1_2_350} that the point-to-point mapping approach is unable to map coarse-grid resolved velocities to SGS stresses and it calculates completely wrong stresses. For the DSM, the PDF shape is similar to the true PDF despite having low cross-correlation. The DSM captures the bulk eddy viscosity, but the phase is completely distorted with the DSM. From Figure~\ref{fig:ts_dnn_di_1_6_350} and Figure~\ref{fig:ts_dnn_di_1_10_350}, we observe an improvement in the prediction of SGS stresses as we start including more features like coarse-grid velocity gradients (i.e., model $\mathbb{M}2$) and the Laplacian on coarse-grid velocities (i.e., model $\mathbb{M}3$). We test the point-to-point ANN for subsampled data using 70 and 24 data snapshots. The cross-correlation between true and predicted stress is almost similar to the one with 350 data snapshots. Figure~\ref{fig:ann_p2p_ns} displays the PDF of true stresses and the predicted stresses computed using point-to-point ANN with a different number of snapshots. Therefore, we can conclude that the ANN can be trained with less number of samples without a significant drop in accuracy. However, neural networks are prone to overfit when we use fewer data and the ability of neural networks to approximate on unseen data reduces. There are different methods to prevent overfitting such as data augmentation, regularization, weight decay, and dropout that should be used when less data is available for learning SGS stresses.   

In terms of computational performance, point-to-point mapping requires less training time for learning SGS stresses from resolved flow variables. This approach is particularly attractive for complex or unstructured mesh and has been applied in many studies \cite{zhu2019machine, ling2016reynolds, wu2018physics, gamahara2017searching, wang2017physics}. As illustrated in these works, our analysis with simple input features like resolved velocities and their derivatives also shows that the input features are critical for effective learning of SGS stresses for point-to-point mapping approach. From Table~\ref{tab:p2p_ann}, we can see that the train time does not increase linearly with an increase in the number of input features. The train time for the neural network mainly depends upon its architecture (how deep and wide it is), and the number of training samples. Since we are using the same architecture for all models, we observe that the train time is similar for all cases. In terms of the test time or deployment time, the point-to-point ANN is slightly slower than DSM (around 1.3 times).     

\begin{table}[htbp]
\caption{Cross-correlation between true and predicted SGS stresses, and CPU time for different models with point-to-point mapping for ANN.}
\label{tab:p2p_ann}       
\begin{tabular}{p{0.1\textwidth} p{0.1\textwidth} p{0.1\textwidth} p{0.1\textwidth} p{0.1\textwidth} p{0.12\textwidth} p{0.12\textwidth}}
\hline\noalign{\smallskip}
Model & $N_s$ & $cc(\tau_{11})$ & $cc(\tau_{12})$ & $cc(\tau_{22})$ & Train time & Test time \\
\noalign{\smallskip}\hline\noalign{\smallskip}
DSM & - & 0.011 & -0.008 & 0.011 & - & 0.0095\\
$\mathbb{M}1$ & 350 & 0.043 & -0.001 & 0.044 & 1577.11 & 0.0138\\
$\mathbb{M}2$ & 350 & 0.343 & 0.261 & 0.343 & 1608.95 & 0.0132\\
$\mathbb{M}3$ & 350 & 0.556 & 0.487 & 0.556 & 1642.82 & 0.0127\\
$\mathbb{M}3$ & 70 & 0.555 & 0.481 & 0.555 & 322.08 & 0.0125\\
$\mathbb{M}3$ & 24 & 0.549 & 0.465 & 0.550 & 112.04 & 0.0128\\
\noalign{\smallskip}\hline
\end{tabular}
\end{table}

\subsection{Neighboring Stencil Mapping} \label{sec:localized_stencil}
\begin{figure*}[htbp]
{\includegraphics[width=0.98\textwidth]{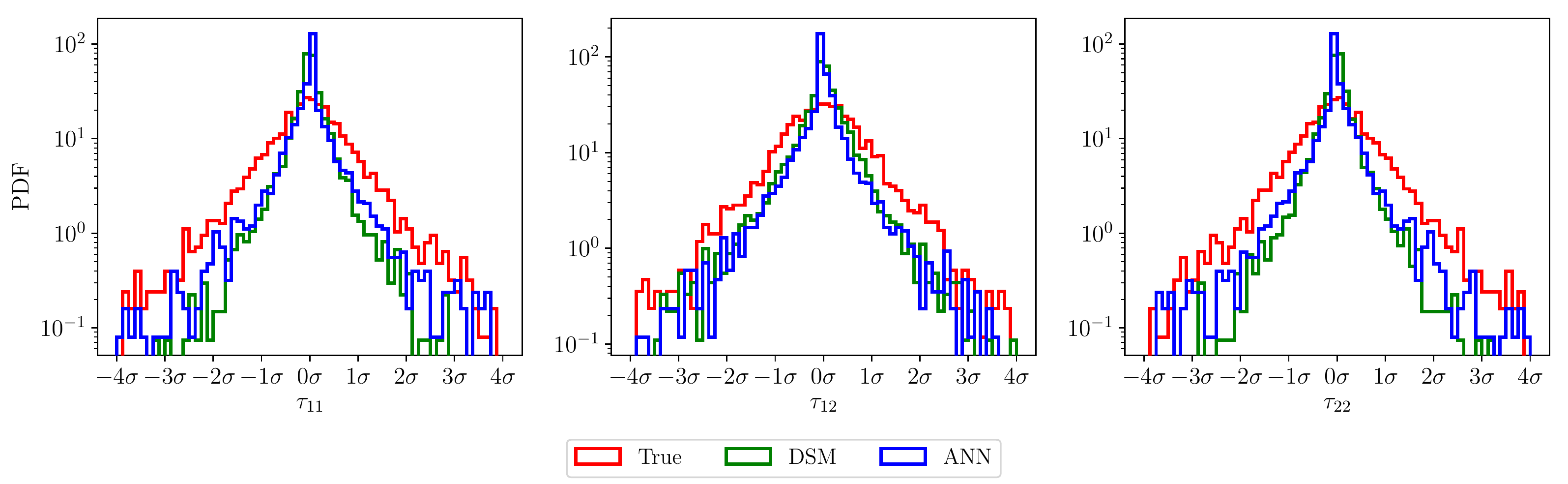}}
\caption{Probability density function for SGS stress distribution with neighboring stencil mapping. The ANN is trained using $\mathbb{M}1 :\{{\bar{u},\bar{v}}\} \rightarrow \{\tilde{\tau}_{11},\tilde{\tau}_{12},\tilde{\tau}_{22}\}$. The training set consists of 350 time snapshots from time $t=0.0$ to $t=3.5$ and the model is tested for $400{th}$ snapshot at $t=4.0$.}
\label{fig:ts_dnn_9_2_350}
\end{figure*}

\begin{figure*}[htbp]
{\includegraphics[width=0.98\textwidth]{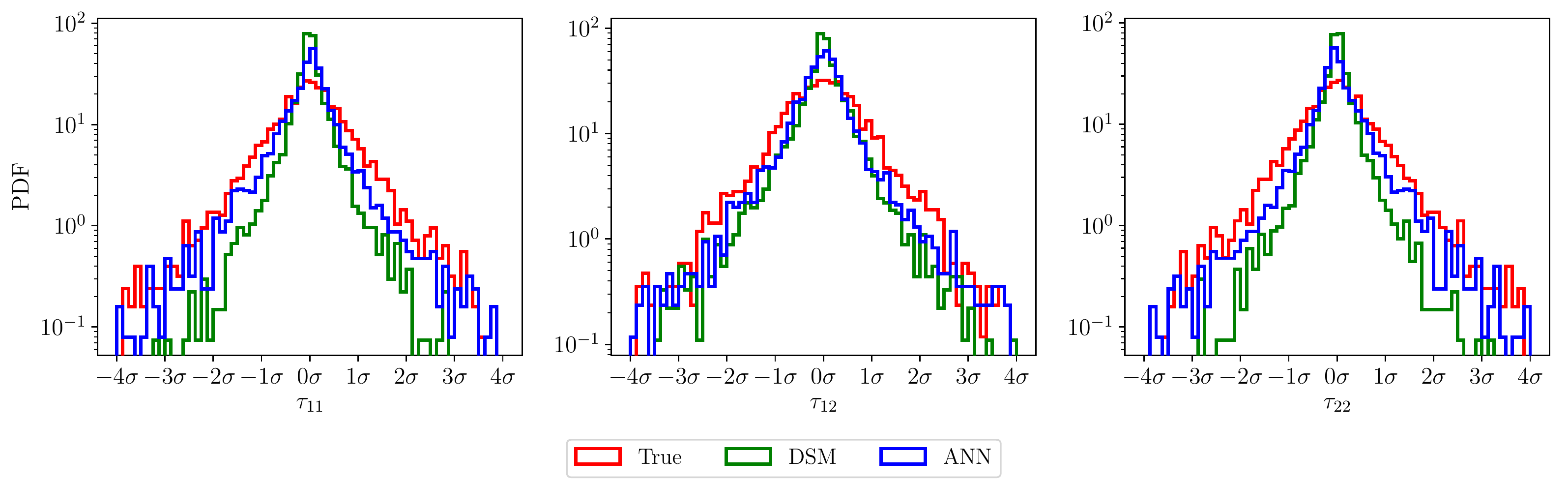}}
\caption{Probability density function for SGS stress distribution with neighboring stencil mapping. The ANN is trained using $\mathbb{M}2 :\{{\bar{u},\bar{v},\bar{u}_x,\bar{u}_x,\bar{u}_y,\bar{v}_y}\} \rightarrow \{\tilde{\tau}_{11},\tilde{\tau}_{12},\tilde{\tau}_{22}\}$. The training set consists of 350 time snapshots from time $t=0.0$ to $t=3.5$ and the model is tested for $400{th}$ snapshot at $t=4.0$.}
\label{fig:ts_dnn_9_6_350}
\end{figure*}

\begin{figure*}[htbp]
{\includegraphics[width=0.98\textwidth]{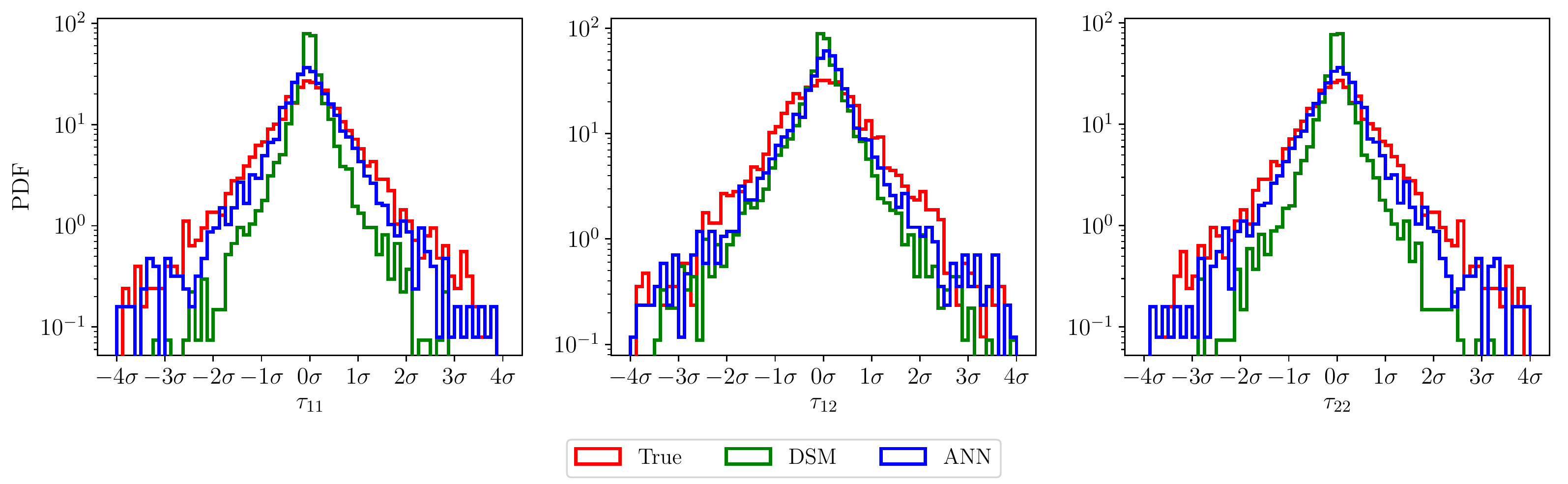}}
\caption{Probability density function for SGS stress distribution with neighboring stencil mapping. The ANN is trained using $\mathbb{M}3 :\{{\bar{u},\bar{v},\bar{u}_x,\bar{u}_x,\bar{u}_y,\bar{v}_y}, \bar{u}_{xx}, \bar{u}_{yy}, \bar{v}_{xx}, \bar{v}_{yy}\} \Rightarrow \{\tilde{\tau}_{11},\tilde{\tau}_{12},\tilde{\tau}_{22}\}$. The training set consists of 350 time snapshots from time $t=0.0$ to $t=3.5$ and the model is tested for $400{th}$ snapshot at $t=4.0$.}
\label{fig:ts_dnn_9_10_350}
\end{figure*}

In this section, we discuss the numerical assessment of results for ANN with neighboring stencil mapping. Table~\ref{tab:lsm_ann} reports the cross-correlation between true and predicted SGS stresses for neighboring stencil mapping ANN. Figure~\ref{fig:ts_dnn_9_2_350} shows that this framework can predict SGS stresses with sufficient accuracy close to the dynamic Smagorinsky model with just coarse-grid velocities (i.e., model $\mathbb{M}1$). If we compare Table~\ref{tab:p2p_ann} and Table~\ref{tab:lsm_ann}, we see that the neighboring stencil mapping with model $\mathbb{M}1$ provides slightly better correlation than utilizing coarse-grid velocities and their derivatives at a single point. This clearly shows the benefit of incorporating neighboring information to determine SGS stresses. As we begin adding more features (i.e., first and second derivative of coarse-grid resolved velocities), we start getting correlation up to 0.8 between true and predicted SGS stresses. From Figure~\ref{fig:ts_dnn_9_6_350} and Figure~\ref{fig:ts_dnn_9_10_350}, we notice that the SGS stresses predicted by ANN are very close to true stresses when the first derivative and Laplacian of coarse-grid velocities are also included in the training. 

\begin{table}[htbp]
\caption{Cross-correlation between true and predicted SGS stresses, and CPU time for different models with neighboring stencil mapping for ANN.}
\label{tab:lsm_ann}       
\begin{tabular}{p{0.1\textwidth} p{0.1\textwidth} p{0.1\textwidth} p{0.1\textwidth} p{0.1\textwidth} p{0.12\textwidth} p{0.12\textwidth}}
\hline\noalign{\smallskip}
Model & $N_s$ &$cc(\tau_{11})$ & $cc(\tau_{12})$ & $cc(\tau_{22})$ & Train time & Test time \\
\noalign{\smallskip}\hline\noalign{\smallskip}
DSM & - & 0.011 & -0.008 & 0.011 & - & 0.0095\\
$\mathbb{M}1$ & 350 & 0.599 & 0.548 & 0.599 & 1675.92 & 0.0136\\
$\mathbb{M}2$ & 350 & 0.783 & 0.731 & 0.783 & 1845.62 & 0.0141\\
$\mathbb{M}3$ & 350 & 0.813 & 0.744 & 0.813 & 2065.11 & 0.0146\\
$\mathbb{M}3$ & 70 & 0.789 & 0.746 & 0.789 & 425.84 & 0.0152\\
$\mathbb{M}3$ & 24 & 0.786 & 0.721 & 0.786 & 139.49 & 0.0149\\
\noalign{\smallskip}\hline
\end{tabular}
\end{table}

We examine this framework with the different number of data snapshots to check the optimal data needed for ANN to learn SGS stresses with sufficient accuracy. From Table~\ref{tab:lsm_ann} we observe that there is a slight drop in cross-correlation as we decrease the amount of data utilized for training. The CPU time required for training drops significantly for less number of training data snapshots. Overall, we can conclude that the accuracy of the prediction will improve with the amount of the training data at the cost of higher computational overhead. One more advantage of this approach is that neighboring stencil mapping can be employed for the complicated and unstructured mesh. This is one of the desirable features of any data-driven frameworks, as the turbulence closure model is deployed for complex fluid flow analysis which is run on supercomputers. In the neighboring stencil mapping framework, the information at only a few neighboring nodes is required and it can be implemented without much of the communication overhead. For the deployment computational time, we get similar findings as to the point-to-point mapping ANN. The neighboring stencil mapping is around 1.5 times slower than DSM. However, to get the same order of accuracy with DSM, we will need to use a fine mesh for LES and this can be computationally expensive than employing neighboring stencil mapping ANN. 

\begin{figure*}[htbp]
{\includegraphics[width=0.98\textwidth]{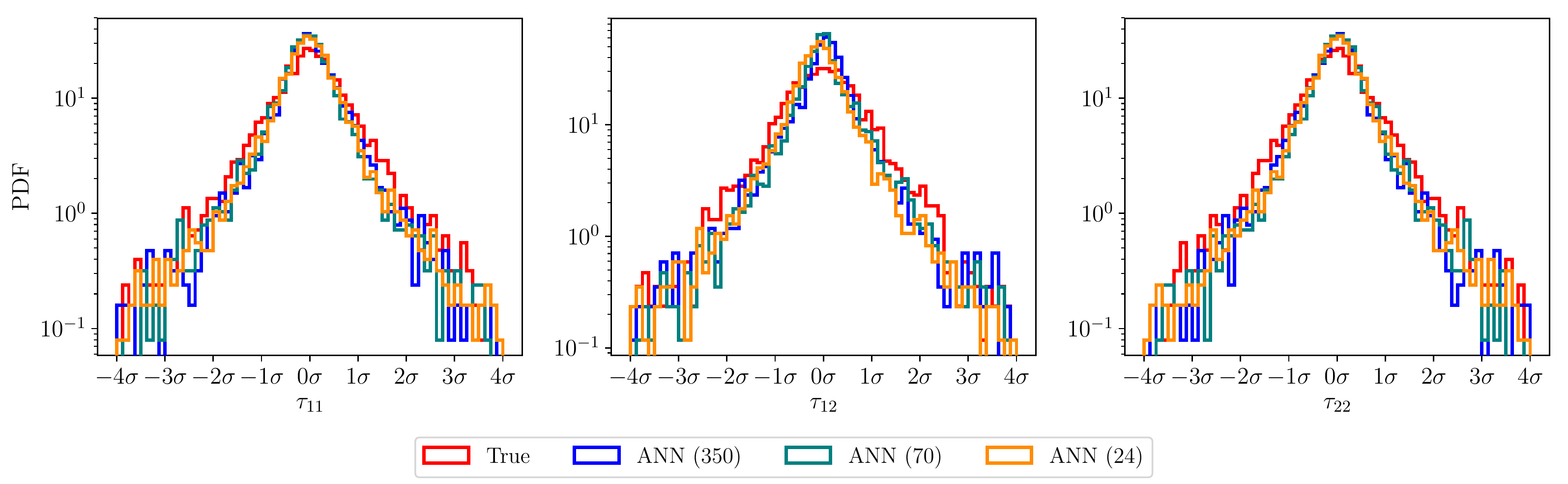}}
\caption{Probability density function for SGS stress distribution with neighboring stencil mapping. The ANN is trained using $\mathbb{M}3 :\{{\bar{u},\bar{v},\bar{u}_x,\bar{u}_x,\bar{u}_y,\bar{v}_y}, \bar{u}_{xx}, \bar{u}_{yy}, \bar{v}_{xx}, \bar{v}_{yy}\} \rightarrow \{\tilde{\tau}_{11},\tilde{\tau}_{12},\tilde{\tau}_{22}\}$. The model is trained using different number of snapshots between $t=0.0$ to $t=3.5$ and the model is tested for $400{th}$ snapshot at $t=4.0$. }
\label{fig:ann_ns}
\end{figure*}

\subsection{CNN Mapping}
\label{sec:cnn_mapping}
\begin{figure*}[htbp]
{\includegraphics[width=0.98\textwidth]{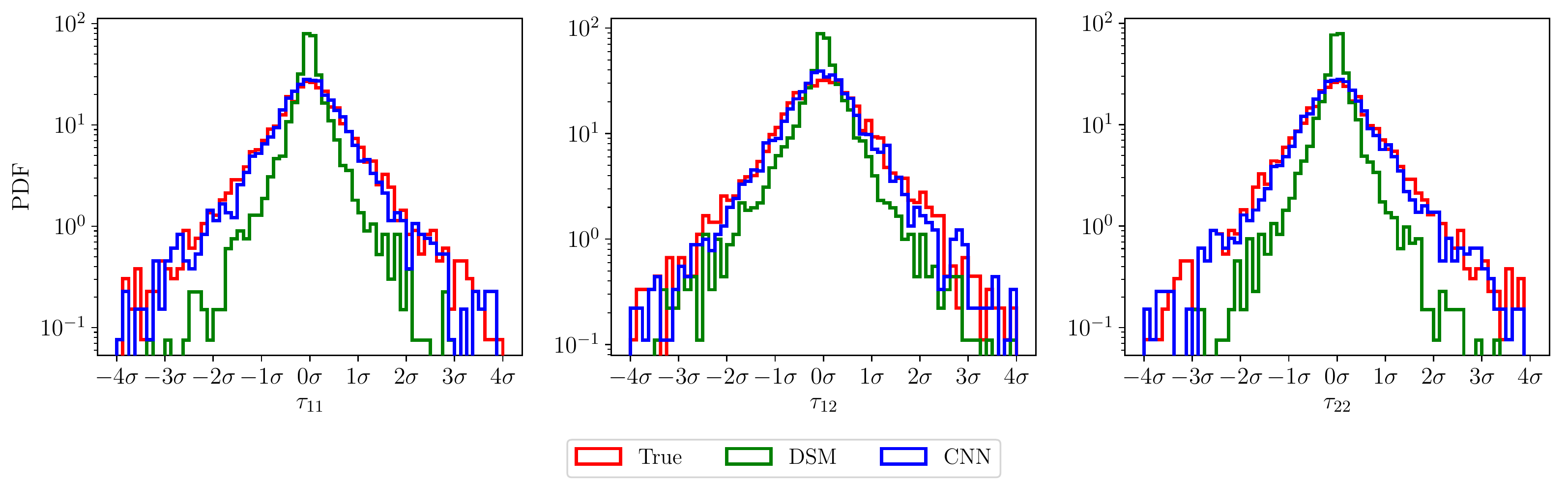}}
\caption{Probability density function for SGS stresses distribution with CNN mapping. The CNN is trained using $\mathbb{M}1 :\{{\bar{u},\bar{v}}\} \rightarrow \{\tilde{\tau}_{11},\tilde{\tau}_{12},\tilde{\tau}_{22}\}$. The training set consists of 350 time snapshots from time $t=0.0$ to $t=3.5$ and the model is tested for $400{th}$ snapshot at $t=4.0$.}
\label{fig:ts_cnn_1_2_350}
\end{figure*}

\begin{figure*}[htbp]
{\includegraphics[width=0.98\textwidth]{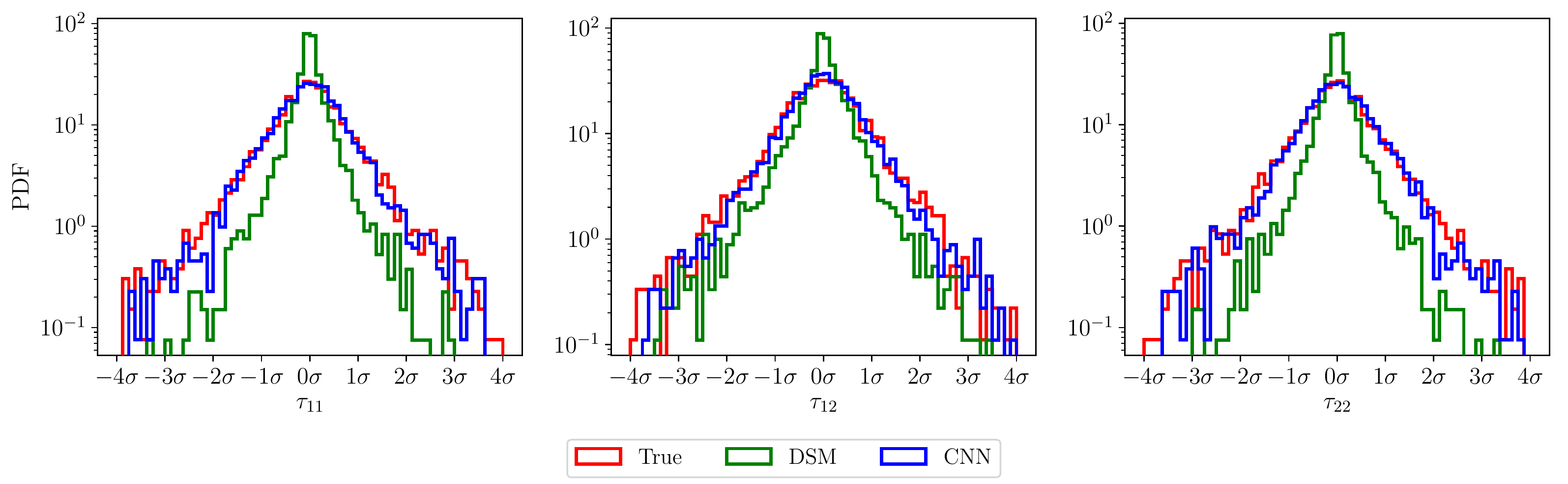}}
\caption{Probability density function for SGS stresses distribution with CNN mapping. The CNN is trained using $\mathbb{M}2 :\{{\bar{u},\bar{v},\bar{u}_x,\bar{u}_x,\bar{u}_y,\bar{v}_y}\} \rightarrow \{\tilde{\tau}_{11},\tilde{\tau}_{12},\tilde{\tau}_{22}\}$. The training set consists of 350 time snapshots from time $t=0.0$ to $t=3.5$ and the model is tested for $400{th}$ snapshot at $t=4.0$.}
\label{fig:ts_cnn_1_6_350}
\end{figure*}

\begin{figure*}[htbp]
{\includegraphics[width=0.98\textwidth]{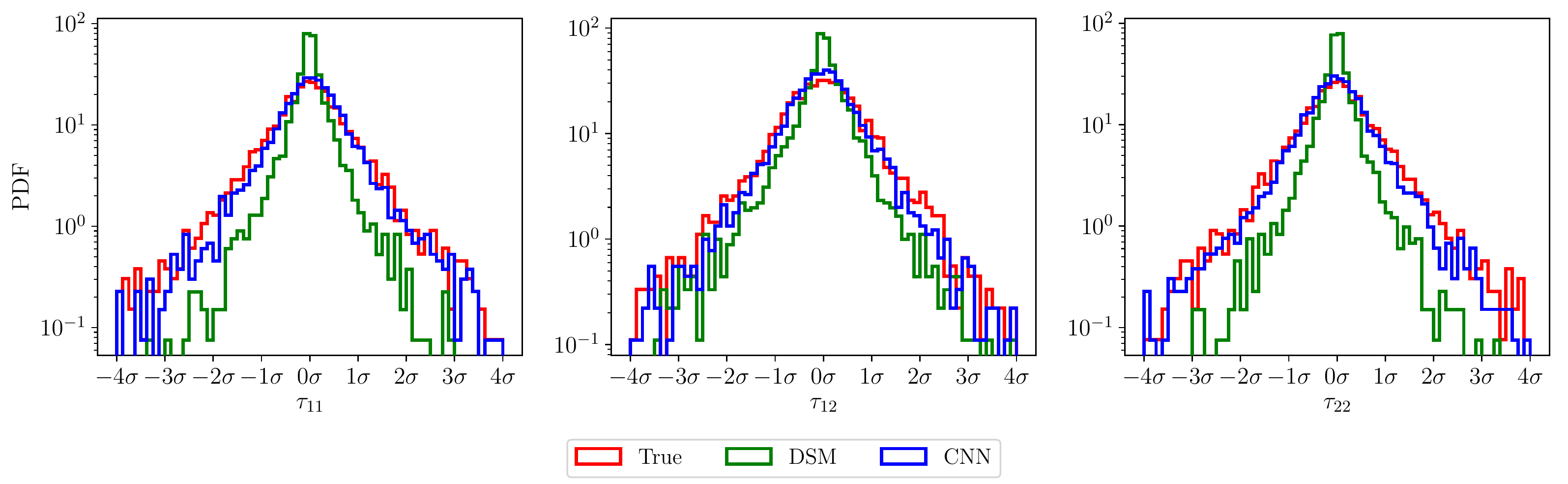}}
\caption{Probability density function for SGS stresses distribution with CNN mapping. The CNN is trained using $\mathbb{M}3 :\{{\bar{u},\bar{v},\bar{u}_x,\bar{u}_x,\bar{u}_y,\bar{v}_y}, \bar{u}_{xx}, \bar{u}_{yy}, \bar{v}_{xx}, \bar{v}_{yy}\} \rightarrow \{\tilde{\tau}_{11},\tilde{\tau}_{12},\tilde{\tau}_{22}\}$. The training set consists of 350 time snapshots from time $t=0.0$ to $t=3.5$ and the model is tested for $400{th}$ snapshot at $t=4.0$.}
\label{fig:ts_cnn_1_10_350}
\end{figure*}

In this section, we present the predictive performance of CNN mapping to learn SGS stresses. Table~\ref{tab:cnn_mapping} lists the cross-correlation between true and predicted SGS stresses computed using CNN mapping. CNN mapping provides the best prediction among three frameworks, and even with just coarse-grid resolved velocities as input features, we obtain cross-correlation around 0.78 between true and predicted SGS stresses. Figure~\ref{fig:ts_cnn_1_2_350} shows the PDF of true and predicted stresses calculated using the model $\mathbb{M}1$ with CNN mapping. CNN mapping can predict the spatial distribution of stresses correctly and we observe that the true and predicted PDF are very close to each other. When we incorporate more input features in the form of first and second derivatives of coarse-grid velocities (i.e., model $\mathbb{M}2$, and $\mathbb{M}3$), we see an improvement in cross-correlation to around 0.84. Figure~\ref{fig:ts_cnn_1_6_350} and Figure~\ref{fig:ts_cnn_1_10_350} display the PDF of true and predicted stresses for models $\mathbb{M}2$ and $\mathbb{M}3$. A very good agreement between PDF of true and predicted SGS stresses are observed for CNN mapping with $\mathbb{M}2$ and $\mathbb{M}3$.

\begin{table}[htbp]
\caption{Cross-correlation between true and predicted SGS stresses, and CPU time for different models with CNN mapping.}
\label{tab:cnn_mapping}       
\begin{tabular}{p{0.1\textwidth} p{0.1\textwidth} p{0.1\textwidth} p{0.1\textwidth} p{0.1\textwidth} p{0.12\textwidth} p{0.12\textwidth}}
\hline\noalign{\smallskip}
Model & $N_s$ &$cc(\tau_{11})$ & $cc(\tau_{12})$ & $cc(\tau_{22})$ & Train time & Test time \\
\noalign{\smallskip}\hline\noalign{\smallskip}
DSM & - & 0.011 & -0.008 & 0.011 & - & 0.0095\\
$\mathbb{M}1$ & 350 & 0.783 & 0.728 & 0.784 & 374.47 & 0.0024\\
$\mathbb{M}2$ & 350 & 0.828 & 0.779 & 0.827 & 391.33 & 0.0021\\
$\mathbb{M}3$ & 350 & 0.835 & 0.779 & 0.835 & 408.65 & 0.0017\\
$\mathbb{M}3$ & 70 & 0.736 & 0.674 & 0.739 & 77.25 & 0.0025\\
$\mathbb{M}3$ & 24 & 0.627 & 0.589 & 0.621 & 27.67 & 0.0025\\
\noalign{\smallskip}\hline
\end{tabular}
\end{table}

\begin{figure*}[htbp]
{\includegraphics[width=0.98\textwidth]{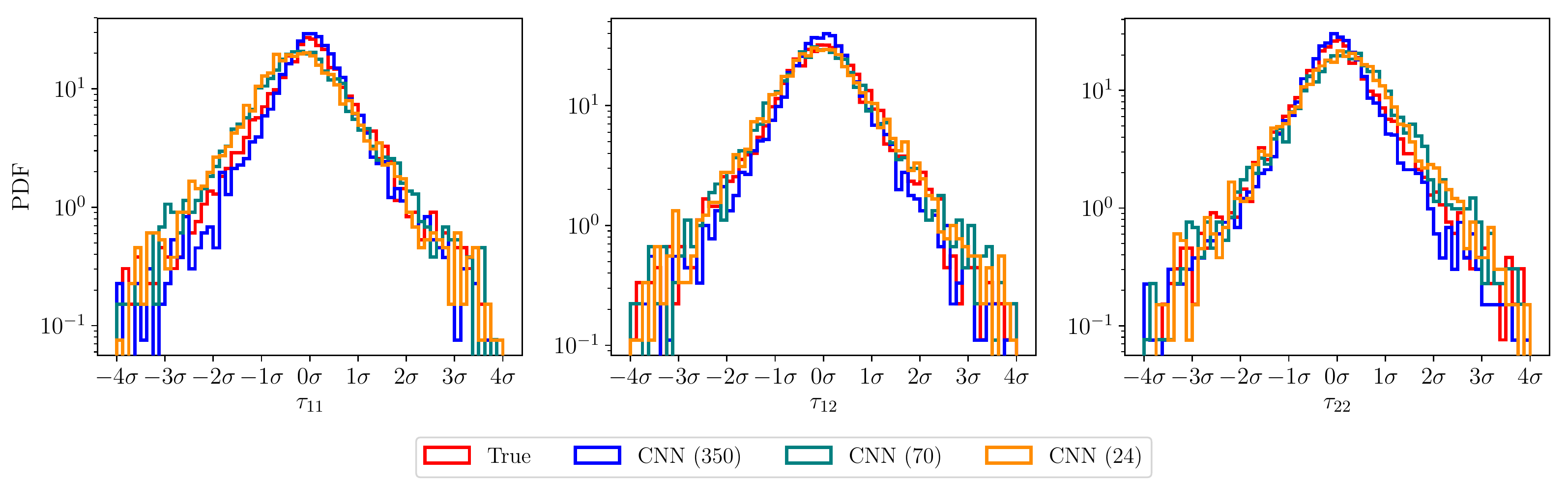}}
\caption{Probability density function for SGS stresses distribution with CNN mapping. The CNN is trained using $\mathbb{M}3 :\{{\bar{u},\bar{v},\bar{u}_x,\bar{u}_x,\bar{u}_y,\bar{v}_y}, \bar{u}_{xx}, \bar{u}_{yy}, \bar{v}_{xx}, \bar{v}_{yy}\} \rightarrow \{\tilde{\tau}_{11},\tilde{\tau}_{12},\tilde{\tau}_{22}\}$. The model is trained using different number of snapshots between $t=0.0$ to $t=3.5$ and the model is tested for $400{th}$ snapshot at $t=4.0$. }
\label{fig:cnn_ns}
\end{figure*}

We also evaluate the performance of CNN mapping with different amount of training snapshots for model $\mathbb{M}3$. Figure~\ref{fig:cnn_ns} shows the PDF of true and predicted stresses for the different number of snapshots. We can see that there is a shift in predicted PDF compared to true PDF for $\tau_{11}$ and $\tau_{22}$ when we use less number of training snapshots. Also the cross-correlation between true and predicted stresses have reduced when we use less number of data snapshots for training, and the performance is poorer than neighboring stencil mapping ANN with less number of snapshots. In terms of computational performance, CNN mapping surpasses both point-to-point and neighboring stencil mapping ANN. This is due to the weight sharing features of CNN and hence the number of parameters to be learned are less than ANN. The deployment computational time for CNN is around 0.2 times the time required by the DSM. Therefore, CNN can provide a more accurate prediction for LES at a less computational cost. Despite these advantages, the application of CNN for the unstructured grid is an open question. If the computational domain has a simple geometry and the data is available in the form of snapshots as in the case of box turbulence, wall-bounded flows, it is advantageous to use CNN. There have been several studies that introduce novel CNN architectures for point cloud data (as in the case of the unstructured grid) \cite{xu2018spidercnn, trask2019gmls, thomas2019kpconv, fey2018splinecnn}. With these novel CNN architectures, the improved predictive capability of CNN can be exploited for turbulence closure modeling. 

\begin{figure*}[htbp]
{\includegraphics[width=0.95\textwidth]{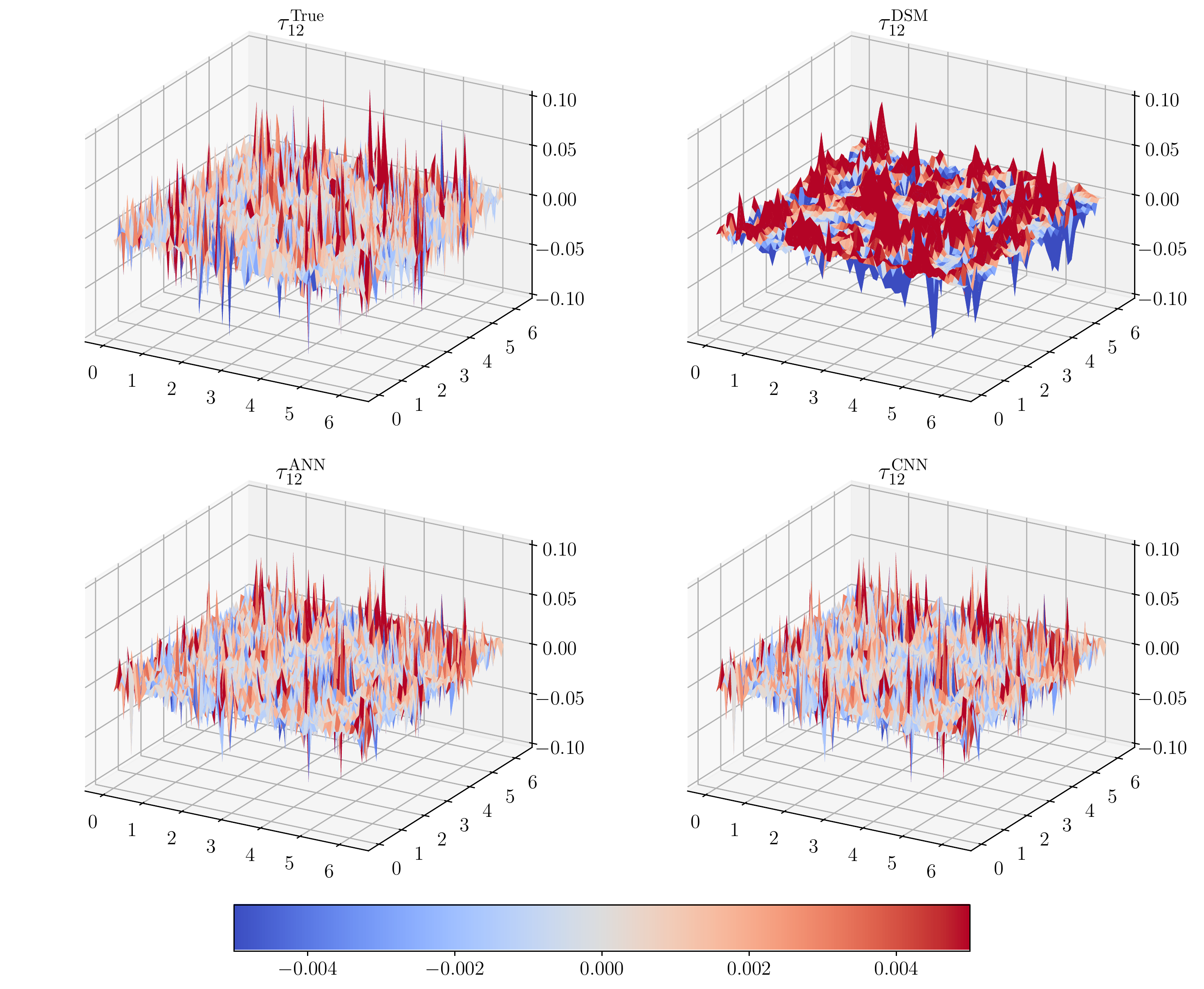}}
\caption{Three-dimensional view of the SGS stress $\tau_{12}$ at time $t=4.0$ by plotting the magnitude along the $z-$axis. $\tau_{12}^{\text{ANN}}$ is the SGS stress computed using neighboring stencil mapping ANN with model $\mathbb{M}3$. $\tau_{12}^{\text{CNN}}$ is the SGS stress computed using CNN mapping with model $\mathbb{M}3$.}
\label{fig:3d_t12}
\end{figure*}

Figure~\ref{fig:3d_t12} displays the three-dimensional view of true SGS stress $\tau_{12}$ and SGS stress predicted by the DSM, neighboring stencil mapping ANN, and CNN mapping. The DSM model captures the bulk eddy viscosity, but not the actual phase. This is the reason behind low value of cross-correlation $cc$ between true SGS stresses and SGS stresses predicted by the DSM. Data-driven models on the other hand are able to capture both magnitude and phase correctly in comparison with true SGS stress $\tau_{12}$.

To summarize our analysis, we show the cross-correlation between true and predicted SGS stresses for all three data-driven closure models with a different number of snapshots. We have summarized the results only for model $\mathbb{M}3$ which includes coarse-grid velocities, coarse-grid velocity gradient, and the Laplacian of coarse-grid velocities. The model $\mathbb{M}3$ was found to give a better prediction for all data-driven models without incurring a high computational cost. It can be clearly seen that CNN are more sensitive to the amount of training data than ANN in terms of its ability to predict SGS stresses. In terms of computational performance, the CNN mapping has the fastest performance (both training and testing/deployment) and has a potential to give accurate prediction with less computational price compared to the dynamic Smagorinsky model.    

\begin{figure*}[htbp]
{\includegraphics[width=0.95\textwidth]{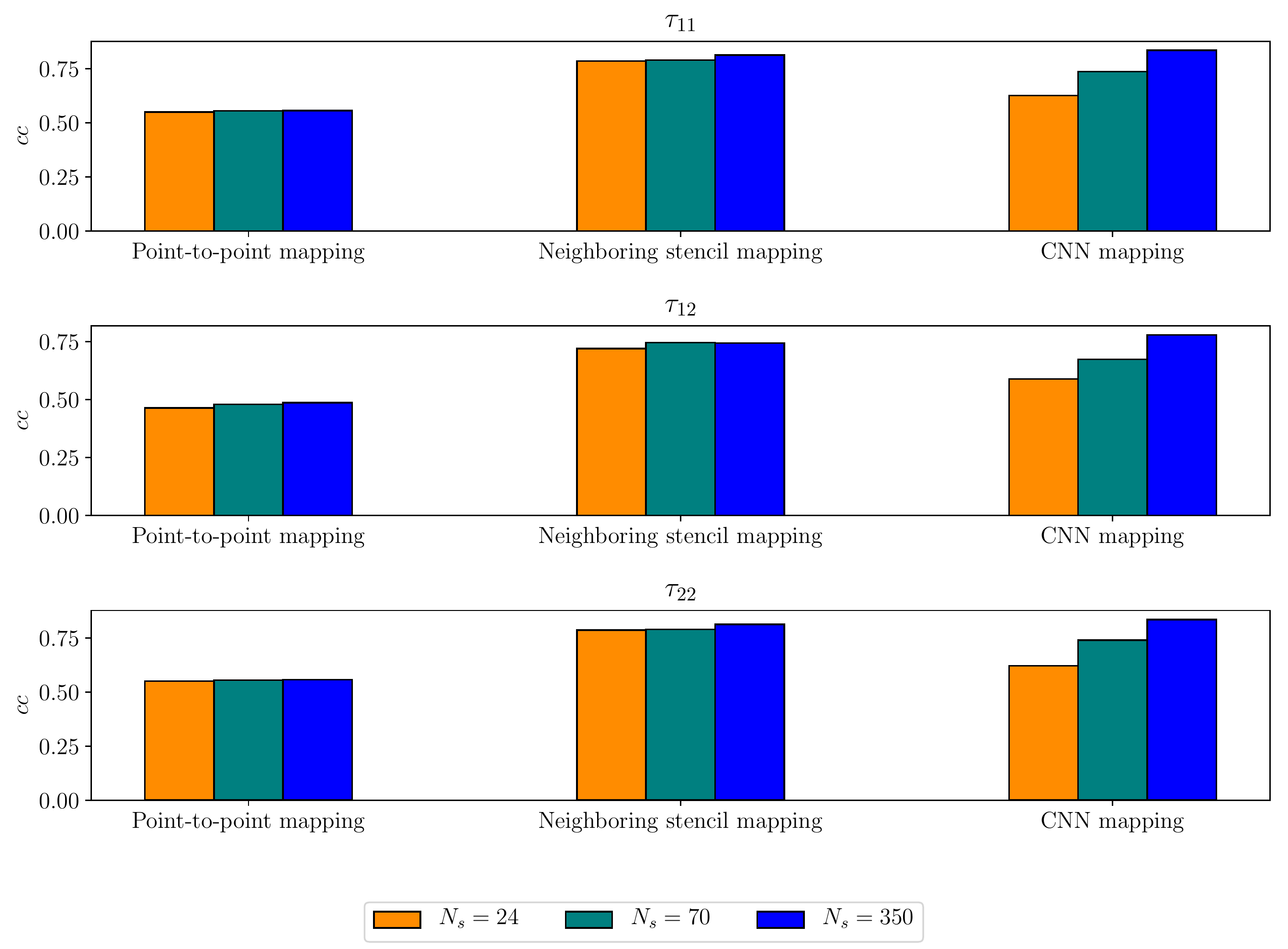}}
\caption{Summary of cross-correlation between true and predicted SGS stresses for different data-driven closure models trained using different number of snapshots.}
\label{fig:hist_summary}
\end{figure*}

\section{Intelligent Eddy Viscosity Modeling}
\label{sec:intelligent_nu}
The data-driven frameworks presented in Section~\ref{sec:numerical_experiments} learn SGS stress directly and hence attempt to improve it's prediction by trying to approximate true SGS stresses. Despite the improved prediction, neural networks are black-box models and these models cannot be interpreted or explained. In this section, we demonstrate intelligent eddy viscosity model as an alternative to the dynamic Smagorinksy model. Our aim here is to illustrate that these black-box data-driven tools can be also tailored to accelerate such phenomenological eddy viscosity models.   

In two-dimensional simulations, the dynamic procedure in the computation of Smagorisnky coefficient in DSM involves the application of low-pass filter eight times at each query. Instead of using these filtering operations, the neural networks can be trained to learn the dynamic eddy viscosity and the trained model can be deployed cost-effectively. One more advantage of this approach is that numerical stability during the \textit{a posteriori} deployment will be enforced. Maulik et al.\cite{maulik2019subgrid} noted that clipping of vorticity source term is required to attain the numerical stability during the deployment of data-driven SGS model. The similar observation was also found by Beck et al. \cite{beck2019deep} for the decaying homogeneous isotropic turbulence problem. The data-driven SGS closure models can predict negative source term at some spatial locations and therefore violates the Boussinesq hypothesis for functional SGS modeling. The intelligent model to learn eddy viscosity can be built by enforcing the constraint such that eddy-viscosity predicted by neural network remains non-negative. This eddy viscosity is then used for computing SGS stresses using Equation~\ref{eq:stat_smag}. We only use coarse-grid velocity and their gradient in the dynamic procedure to compute the Smagorinsky coefficient. Hence we can include them as input features to learn eddy viscosity. The intelligent eddy viscosity model is given as  
\begin{equation}
    \mathbb{M}4 : \{\bar{u}, \bar{v}, \bar{u}_x, \bar{u}_y, \bar{v}_x, \bar{v}_y\} \in \mathbb{R}^6 \rightarrow \{ \nu_e\} \in \mathbb{R}^1, \label{eq:m4}
\end{equation}
where the eddy viscosity $\nu_e$ is given as 
\begin{equation}
    \nu_e = (C_s \Delta)^2|\bar{S}|,
\end{equation}
where $(C_s \Delta)^2$ is computed from Equation~\ref{eq:cs2}, and $|\bar{S}|$ is given by Equation~\ref{eq:S}. The similar framework was studied by Pal \cite{pal2019deep}, and they showed that the data-driven model gives two to eight times computational performance gain against the dynamic Smagorinsky model for wall-bounded turbulent flows.

\begin{figure*}[htbp]
{\includegraphics[width=0.98\textwidth]{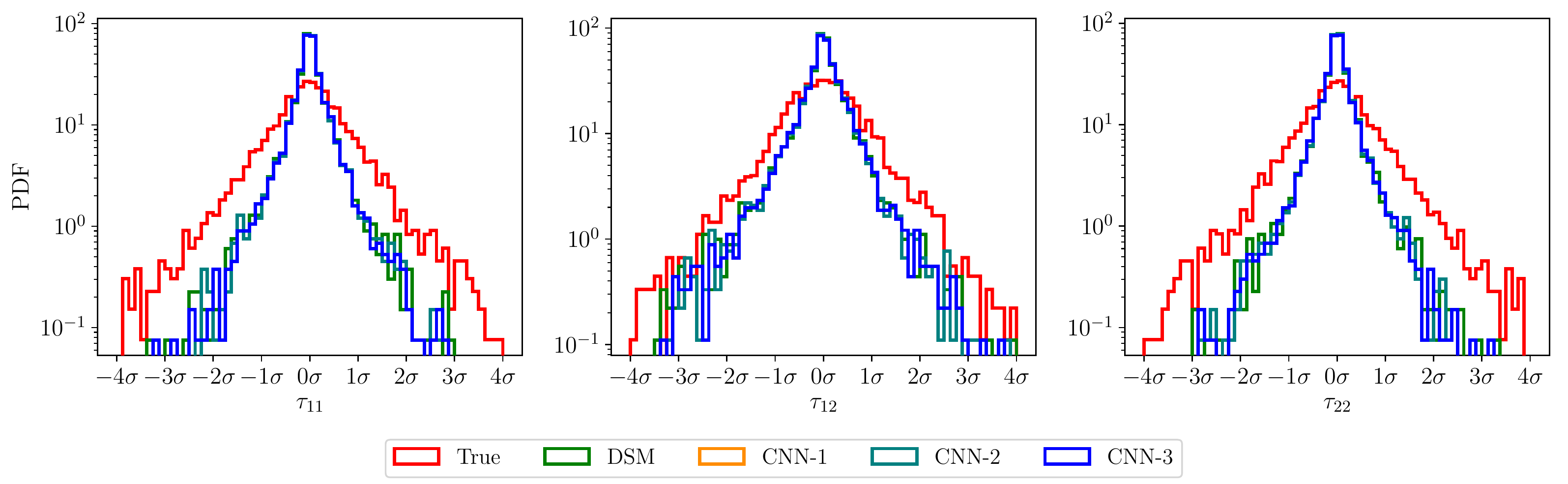}}
\caption{Probability density function for true SGS stresses distribution and stresses computed at $t=4.0$ with DSM and intelligent eddy viscosity model. The CNN is trained using $\mathbb{M}4. :\{{\bar{u},\bar{v},\bar{u}_x,\bar{u}_x,\bar{u}_y,\bar{v}_y}\} \rightarrow \{\nu_e\}$. The model is trained using 350 snapshots between $t=0.0$ to $t=3.5$ and the prediction is shown for time $t=4.0$. }
\label{fig:cnn_nu}
\end{figure*}

The main advantage of this modeling approach is the numerical stability during \textit{a posteriori} deployment and computational speed up. To avoid repetition, we compare the performance of intelligent eddy viscosity model with CNN mapping only.  Table~\ref{tab:cnn_inu} lists the performance of intelligent eddy viscosity model trained with a different number of hyperparameters. The task of learning dynamic eddy viscosity is easier as compared to learning true SGS stresses and hence, we get cross-correlation up to 0.98 with just one hidden layer and 16 kernels. The intelligent eddy viscosity model is around eight times faster than the DSM. If we use deep network similar to the data-driven SGS model, we still get a computational speedup of four times. Figure~\ref{fig:cnn_nu} show the comparison of true SGS stresses, SGS stresses predicted by DSM, and SGS stresses computed from different intelligent eddy viscosity models (different CNN architectures). The SGS stresses predicted by CNN are very close to the stresses computed from the DSM, and hence we can get similar performance similar to the DSM at much less computational cost. 

\begin{table}[htbp]
\caption{Cross-correlation between DSM eddy viscosity and intelligent eddy viscosity predicted by data-driven models, and CPU time for different models with CNN mapping.}
\label{tab:cnn_inu}       
\begin{tabular}{p{0.1\textwidth} p{0.2\textwidth} p{0.1\textwidth} p{0.12\textwidth} }
\hline\noalign{\smallskip}
Model & Hyperparameters & $cc(\nu_e)$ & Test time \\
\noalign{\smallskip}\hline\noalign{\smallskip}
DSM & - & - & 0.0095\\
CNN-1 & [16] & 0.988 & 0.0012 \\
CNN-2 & [16,8,16] & 0.994 & 0.0015 \\
CNN-3 & [16,8,8,8,8,16] & 0.992 & 0.0024 \\
\noalign{\smallskip}\hline
\end{tabular}
\end{table}

\section{Conclusion}
\label{sec:conclusion}
In the present study, we investigated different data-driven turbulence closure frameworks to learn SGS stresses using coarse-grained field variables. The blending of data-driven turbulence closure models within physics-based LES framework present the hybrid modeling approach that has the potential to give accurate prediction of fluid flows at a less computational cost. The traditional Smagorinsky model is based on empirical formulas and phenomenological relationships and can either produce insufficient or excessive dissipation. On the other hand, the optimal map between coarse-grid field variables and SGS stresses learned by data-driven framework provides improved prediction compared to the dynamic Smagorinsky model. The importance of the selection of input features in the prediction of SGS stresses is illustrated for different data-driven closure models using two-dimensional Kraichnan turbulence as the prototype example. The quantitative analysis using cross-correlation indicates that the prediction of SGS stresses improves when coarse-grid velocities. velocity gradients and their Laplacian are included in input features. The analysis with localized mapping showed that the improvement in the prediction of SGS stresses is achieved when information from neighboring points is also included without any significant increase in training and deployment time. The CNN mapping provides the most accurate prediction close to true SGS stresses with less computational overhead for training because of their invariance and weight sharing property.       

The analysis of deployment time for different frameworks points out that data-driven closure models can give accurate SGS stresses prediction with the same or less computational overhead as the dynamic Smagorisnky model. The localized point-to-point mapping with ANN is particularly attractive for practical engineering applications due to its ability to handle unstructured mesh. The CNN mapping, on the other hand, seems more suitable for applications where a large amount of training data is available in the form of snapshots. While intelligent SGS modeling frameworks can model true SGS stresses accurately, they are prone to numerically unstable prediction in the \textit{a posteriori} deployment as shown in recent studies \cite{maulik2019subgrid, beck2019deep}. To exploit the potential of these black-box models in safety critical applications, we further investigate their robustness in predicting eddy viscosity coefficient. Although limiting their predictive accuracy by utilizing an eddy viscosity model, we illustrate that the intelligent eddy viscosity approach gives four to eight times computational speedup with the same accuracy as the DSM.    

We highlight that the data-driven techniques are in their infancy. Once the trained model is deployed in a CFD code, \textit{a posteriori} analysis of data-driven closure models might give unexpected predictions (i.e., numerically or physically inconsistent). To be able to use the neural network based model in safety-critical applications, we need either of the two: interpret the model and figure out when it can fail, or to use neural networks in a way that we can detect when it fails and produces nonphysical results. Interpreting a deep neural network with millions of parameters is almost impossible. In our future work, we will focus on the second approach with internal sanity checking mechanism where a black box model helps better modeling of conservation laws, and conservation mechanism puts a sanity check on black-box model. Furthermore, the neural-network architectures employed in this work are fairly simple plain vanilla versions without any complex structure. The predictive performance of data-driven closure models can be further improved by constructing more sophisticated architecture designs like TBNN \cite{ling2016reynolds}, and generative adversarial networks \cite{goodfellow2014generative}. In the future, we would also like to extend these approaches for more complex test cases such as three-dimensional Kolmogorov turbulence, and geophysical flows.


\section*{Acknowledgements}
This material is based upon work supported by the U.S. Department of Energy, Office of Science, Office of Advanced Scientific Computing Research under Award Number DE-SC0019290. O.S. gratefully acknowledges their support. 

Disclaimer: This report was prepared as an account of work sponsored by an agency of the United States Government. Neither the United States Government nor any agency thereof, nor any of their employees, makes any warranty, express or implied, or assumes any legal liability or responsibility for the accuracy, completeness, or usefulness of any information, apparatus, product, or process disclosed, or represents that its use would not infringe privately owned rights. Reference herein to any specific commercial product, process, or service by trade name, trademark, manufacturer, or otherwise does not necessarily constitute or imply its endorsement, recommendation, or favoring by the United States Government or any agency thereof. The views and opinions of authors expressed herein do not necessarily state or reflect those of the United States Government or any agency thereof.

\section*{Appendix A: Derivation of the Smagorinsky model in 2D turbulence}
\label{sec:smag}
From Equation~\ref{eq:sgs}, the subgrid-scale stresses in 2D field can be written as 
\begin{align}
    \tau_{ij} &= \overline{u_i u_j} -  \bar{u}_i \bar{u}_j, \\
              &= \underbrace{\frac{1}{2}\tau_{kk}\delta_{ij}}_{k_{\text{SGS}}\delta_{ij}} + \bigg(\underbrace{ \tau_{ij} - \frac{1}{2}\tau_{kk}\delta_{ij}}_{\tau_{ij}^d}  \bigg).
\end{align}
The SGS stresses can be written as
\begin{equation}\label{eq:ksgs}
    \tau = k_{\text{SGS}}I + \tau^d,
\end{equation}
where $k_{\text{SGS}}=\frac{1}{2}\tau_{kk}$ is called subgrid-scale kinetic energy (i.e., using the conventional summation notation with repeating indices, for example, $\tau_{kk} = \tau_{11} + \tau_{22}$, in 2D). In Smagorinsky model, we model the deviotoric (traceless) part of SGS stresses as 
\begin{equation}\label{eq:modelM}
    \tau_{ij}^d = -2\nu_e\bar{S}_{ij}^d,
\end{equation}
where $\nu_e$ is the SGS eddy viscosity, and $\bar{S}_{ij}$ is called resolved strain rate tensor given by
\begin{equation}
    \bar{S}_{ij} = \frac{1}{2}\bigg( \frac{\partial \bar{u}_i}{\partial x_j} + \frac{\partial \bar{u}_j}{\partial x_i} \bigg),
\end{equation}
where we can write explicity as follows
\begin{equation}
    \bar{S} = \begin{bmatrix}
    \frac{\partial \bar{u}}{\partial x}       & \frac{1}{2}\bigg( \frac{\partial \bar{u}}{\partial y} + \frac{\partial \bar{v}}{\partial x} \bigg) \\
    \frac{1}{2}\bigg( \frac{\partial \bar{v}}{\partial x} + \frac{\partial \bar{u}}{\partial y} \bigg)       & \frac{\partial \bar{v}}{\partial y} 
    \end{bmatrix} .
\end{equation}
The trace of the $\bar{S}$ is zero owing to the continuity equation for incompressible flows. Therefore, $\bar{S}_{ij}^d = \bar{S}_{ij}$ and the Smagorinsky model becomes
\begin{equation}\label{eq:smag}
    \tau_{ij}^d = -2\nu_e\bar{S}_{ij}.
\end{equation}
The eddy viscosity approximation computes $\nu_e$ using the below relation
\begin{equation}\label{eq:nu}
    \nu_e = C_k \Delta \sqrt{k_{\text{SGS}}},
\end{equation}
where the proportionality constant is often set to $C_k = 0.094$, and $\Delta$ is the length scale (usually grid size). The SGS kinetic energy $k_{\text{SGS}}$ is computed with the local equilibrium assumption of the balance between subgrid scale energy production and dissipation
\begin{equation}\label{eq:equilibrium}
    \bar{S}:\tau + C_{\epsilon}\frac{k_{\text{SGS}}^{1.5}}{\Delta} = 0,
\end{equation}
where the first term in above equation is dissipation flux, second term is production flux, and the production constant is often set to $C_{\epsilon}=1.048$. The double inner product operation $:$ is given by 
\begin{equation}
    \bar{S}:\tau = \bar{S}_{ij}\tau_{ij}=\bar{S}_{11}\tau_{11} + \bar{S}_{12}\tau_{12} +  \bar{S}_{21}\tau_{21} + \bar{S}_{22}\tau_{22}.    
\end{equation}
Substituting Equation~\ref{eq:ksgs} and \ref{eq:smag} int Equation~\ref{eq:equilibrium}, we get
\begin{align}
    \bar{S}:(k_{\text{SGS}}I - 2C_k \Delta \sqrt{k_{\text{SGS}}}\bar{S}) + C_{\epsilon}\frac{k_{\text{SGS}}^{1.5}}{\Delta} &= 0, \\
    \sqrt{k_{\text{SGS}}}\bigg( \frac{C_{\epsilon}}{\Delta}k_{\text{SGS}} + \sqrt{k_{\text{SGS}}} \underbrace{\bar{S}:I}_{\bar{S}_{ij} \delta_{ij} = 0} -  2C_k \Delta \bar{S}:\bar{S} \bigg) &= 0, \\
    \frac{C_{\epsilon}}{\Delta}k_{\text{SGS}} - 2C_k \Delta \bar{S}:\bar{S} &= 0, 
\end{align}
From above equations, subgrid-scale kinetic energy can be written as
\begin{align}
    k_{\text{SGS}} &= \frac{C_k}{C_{\epsilon}}\Delta^2(2\bar{S}:\bar{S}), \\
    k_{\text{SGS}} &= \frac{C_k}{C_{\epsilon}}\Delta^2|\bar{S}|^2,
\end{align}
where $|\bar{S}| = \sqrt{2 \bar{S}_{ij} \bar{S}_{ij}}$. Furthermore, substituting Equation~\ref{eq:nu} in above Equation, we get
\begin{equation}
    \nu_e = C_k \Delta^2 \sqrt{\frac{C_k}{C_\epsilon}}|\bar{S}|.
\end{equation}
We can define a new constant coefficient as 
\begin{equation}
    C_s^2 = C_k \sqrt{\frac{C_k}{C_\epsilon}}.
\end{equation}
where $C_s=0.1678$ is called the Smagorinsky coefficient. Finally, we get below expression for SGS eddy viscosity 
\begin{equation}
    \nu_e = C_s^2 \Delta^2 |\bar{S}|,
\end{equation}
and the Smagorinsky model, given by Equation~\ref{eq:modelM}, reads as
\begin{equation}
    \tau_{ij}^{d} = -2C_s^2 \Delta^2 |\bar{S}|\bar{S}_{ij}.
    \label{eq:static_smag}
\end{equation}

\section*{Appendix B: Hyperparameters Optimization}
\label{sec:appendix_hp}
In the appendix, we outline the procedure we followed for selection of hyperparameters for ANN with point-to-point mapping and neighboring stencil mapping. For ANN, there are many hyperparameters such as number of neurons, number of hidden layers, loss function, optimization algorithm, activation function, and batch size etc. If we use regularization, dropout, or weight decay to avoid overfitting, the design space of hyperparameters increases further. 

We focus on three main hyperparameters of ANN: number of neurons, number of hidden layers, and learning rate of optimization algorithm. The training data is scaled between [-1,1] using the minimum and maximum value in the training dataset. We use ReLU activation function which is given by $\zeta(\chi) = \text{max}(0,\chi)$, where $\zeta$ is the activation function, and $\chi$ is the input to the node. We use Adam optimization algorithm \cite{kingma2014adam} and the batch size is kept constant at 256. Adam optimization algorithm has three hyperparameters: learning rate $\alpha$, first moment decay rate $\beta_1$, and second moment decay rate $\beta_2$. We test our ANN for two learning rates $\alpha=0.001$, and $0.0001$. The other two hyperparameters in Adam optimization algorithm are $\beta_1=0.9$, and $\beta_2=0.999$. We employ mean squared error as the loss functions, since it is a regression problem. We test both ANN with point-to-point mapping and neighboring stencil mapping for four different number of hidden layers $L=2,3,5,7$. The ANN with point-to-point mapping is tested for four different number of neurons $N=20,30,40,50$ and the local-stencil-mapping is tested for $N=40,60,80,100$. The number of neurons is higher in case of local-stencil-mapping because there are more features compared to point-to-point mapping. 

The optimal ANN architecture is selected using multi-dimensional gridsearch algorithm coupled with k-fold cross-validation. Cross-validation is a procedure used to determine the performance of the neural network on unseen data. The procedure consists of dividing the training data into k groups, training the ANN by excluding each group and evaluating the model's performance on that group. Therefore, if we use five-fold cross-validation then the model is trained five times and the performance index is computed for five groups. Once the performance for each group is available, the mean of the performance index is utilized to select optimal hyperparameters. We use 500 epochs for determining the optimal hyperparameters. A good learning is achieved when both training loss and validation loss reduce till the learning rate is minimal. We apply coefficient of determination $r^2$ as the performance index to decide optimal hyperparameters. The calculation of coefficient of determination is done using below formula
\begin{equation}
    r^2 = 1 - \frac{\sum_{i}(y_i-\tilde{y}_i)^2}{\sum_{i}(y_i-\bar{y})^2},
\end{equation}
where $y_i$ is the true label, $\tilde{y}$ is the predicated label, and $\bar{y}$ is the mean of true labels.

\begin{figure*}[htbp]
\centering
{\includegraphics[width=0.9\textwidth]{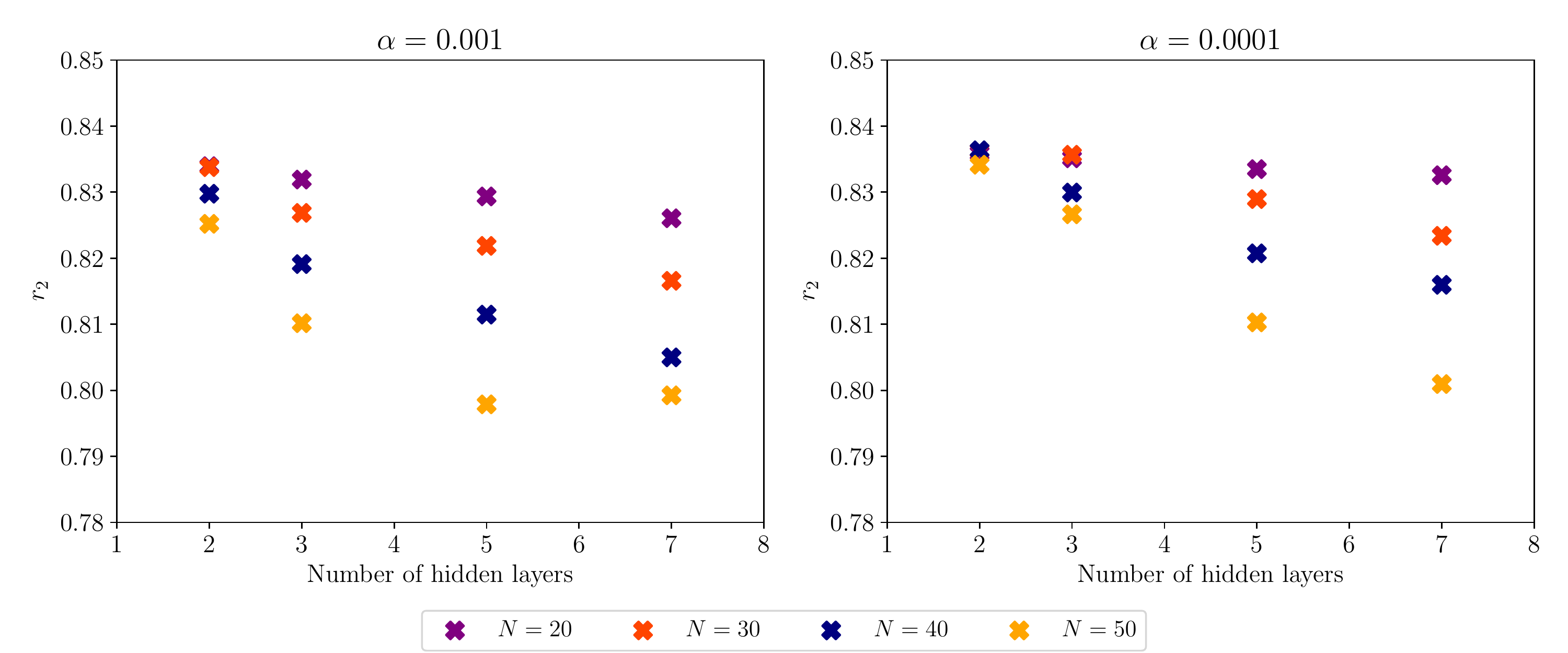}}
\caption{Hyperparameters search using the gridsearch algorithm combined with five-fold cross validation for the neural network using point-to-point mapping with $\mathbb{M}3$.}
\label{fig:hp_direct_10}
\end{figure*}

\begin{figure*}[htbp]
\centering
{\includegraphics[width=0.9\textwidth]{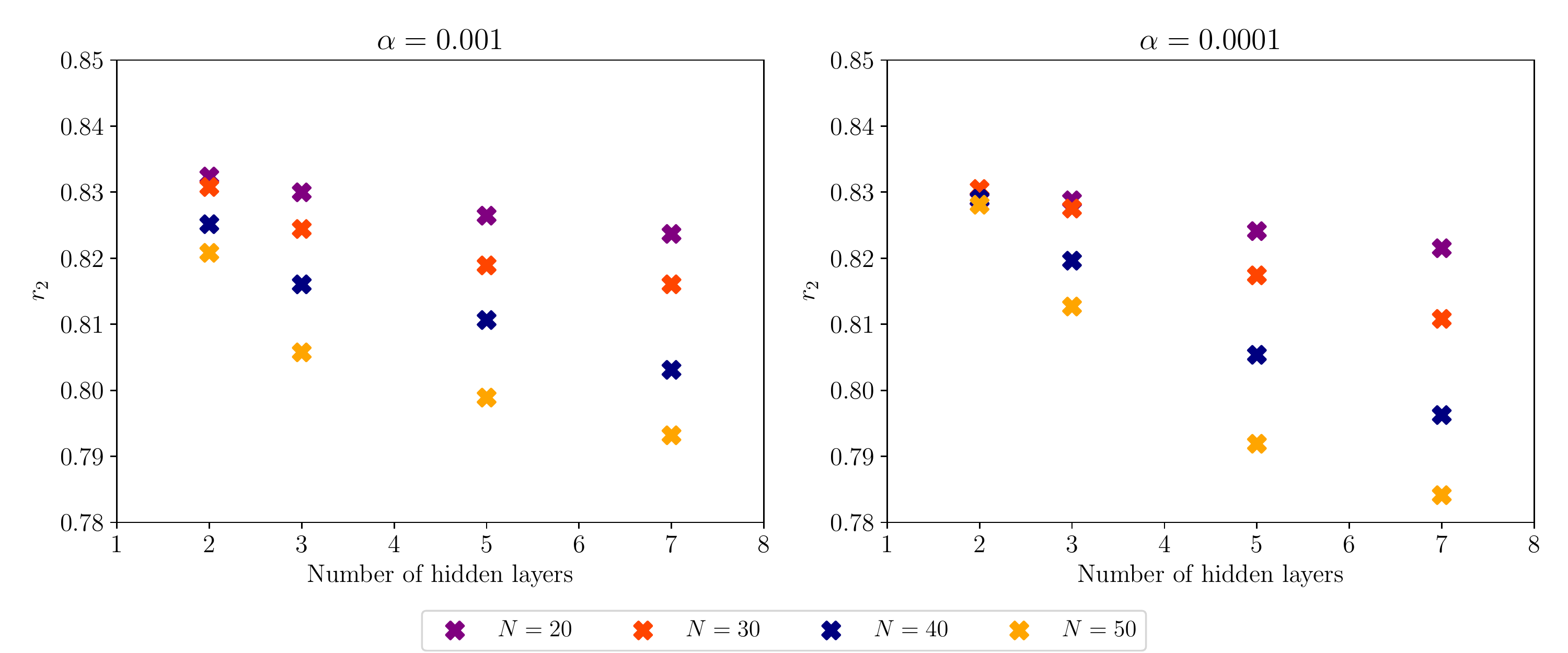}}
\caption{Hyperparameters search using the gridsearch algorithm combined with five-fold cross validation for the neural network using neighboring stencil mapping with $\mathbb{M}3$.}
\label{fig:hp_direct_9_10}
\end{figure*}

Figure~\ref{fig:hp_direct_10} displays the performance index for ANN with point-to-point mapping and $\mathbb{M}3$ model for all hyperparameters tested using gridsearch algorithm. It can be observed that the performance of the network does not change significantly with hyperparameters and the difference in performance is very small. The optimal hyperparameters obtained for point-to-point mapping ANN are $L=2$, $N=40$, and $\alpha=0.0001$. We use the same hyperparameters for other two models $\mathbb{M}1$ and $\mathbb{M}2$ for point-to-point mapping ANN. We see the similar behaviour in case of neighboring stencil mapping ANN and model $\mathbb{M}3$ as shown in Figure~\ref{fig:hp_direct_9_10}. The optimal hyperparameters for neighboring stencil mapping ANN are $L=2$, $N=40$, and $\alpha=0.001$.

\begin{figure*}[htbp]
\centering
{\includegraphics[width=0.9\textwidth]{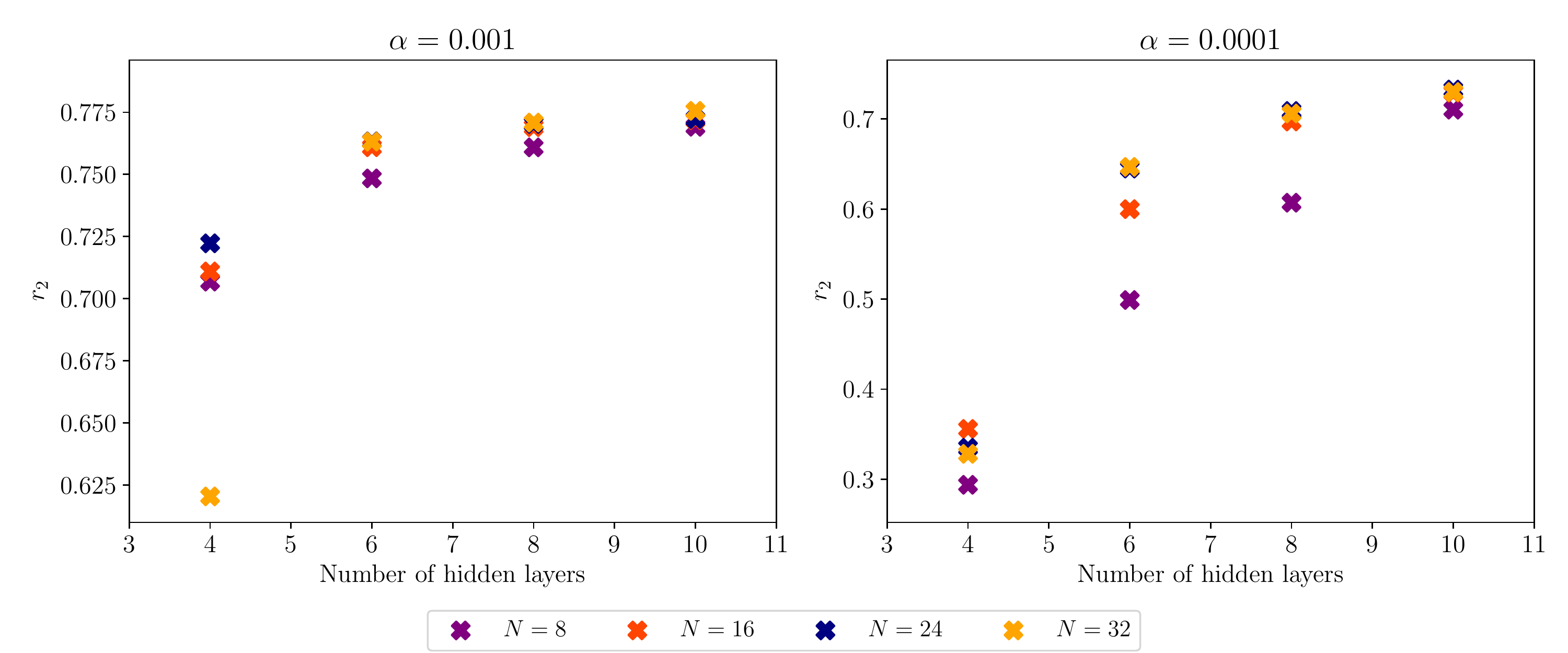}}
\caption{Hyperparameters search using the gridsearch algorithm combined with five-fold cross validation for CNN mapping with model $\mathbb{M}3$.}
\label{fig:hp_cnn}
\end{figure*}

The CNN architecture has similar hyperparameters as the ANN. Additionally, we need to select the kernel shape and strides for CNNs. Stride is the amount by which the kernel should shift as it convolves around the volume. We use the stride=1 in both $x$ and $y$ directions. We use $3 \times 3$ shaped kernel in our CNN architecture. We check the performance of CNN architecture for different number of hidden layers $L=2,4,6,8$, different number of filters $N=8,16,24,32$, and two learning rates. Figure~\ref{fig:hp_cnn} displays the performance index of CNN for different hyperparameters. The performance of CNN is more sensitive to the learning rate and we observe stable performance for the learning rate $\alpha=0.001$. The performance is almost similar for $L=6,8,10$ with different number of kernels. We can select $L=6$ and $N=16$ which has performance index of 0.76. Additionally, we test the CNN architecture with $L=6$ and $[16,8,8,8,8,16]$ distribution for the number of kernels along hidden layers and we observed the performance index of 0.75 at less computational cost. Therefore, we apply $L=6$, $N=[16,8,8,8,8,16]$, and $\alpha=0.001$ as our hyperparameters for the CNN architecture.

\section*{Appendix C: CPU Time Measurements}
\label{sec:appendix_cpu}
In this study, the pseudo-spectral solver used for DNS is written in Python programming language. The code for coarsening of variables from fine to coarse grid, dynamic Smagorinsky model code is all written in Python. We use vectorization to get faster computational performance. The machine learning library Keras is also available in Python and is used for developing all data-driven closure models. Therefore, the CPU time reported in our analysis is for codes which are all developed on the same platform. We would like to highlight that when the trained model is deployed, it makes the function for first time and hence it takes slightly more time. Once the function is created, the CPU time for deployment is less. Therefore, in all our Tables, we report the CPU time for running the predict function second time since initializing CUDA kernels might yield a startup overhead as shown in Listing~\ref{lst:cpu}, where t1 here has some idle time due to initializing kernels. In our study, we report t2, and we further verified that t3 - t2 = t2, which illustrate that the reported CPU times are consistent.  

\begin{lstlisting}[frame=single,language=Python, caption=Code sample to check the CPU time for data-driven models. ,label = {lst:cpu} ]
test_time_init = tm.time()
y_test = model.predict(ftest)
t1 = tm.time() - test_time_init
       
test_time_init = tm.time()
y_test = model.predict(ftest)
t2 = tm.time() - test_time_init

test_time_init = tm.time()
y_test = model.predict(ftest)
y_test = model.predict(ftest)
t3 = tm.time() - test_time_init 
\end{lstlisting}

\section*{Appendix D: ANN and CNN architectures}
\label{sec:appendix_codes}
We use open-source Keras library to build our neural networks. It uses TensorFlow at the backend. Keras is widely used for fast prototyping, advanced research and production due to its simplicity and faster learning rate. Keras library provides different options for optimizers, neural network architectures, activation functions, regularization, dropuout, etc. Any simple neural network architecture can be coded with few lines of code. The sample code for ANN and CNN used in this work are listed in Listings~\ref{lst:ann} and \ref{lst:cnn}. 

\begin{lstlisting}[frame=single, language=Python, caption=Sample code for the ANN used in this study. ,label = {lst:ann}]
model = Sequential()

input_layer = Input(shape=(nf,))

x = Dense(40, activation='relu',  use_bias=True)(input_layer)
x = Dense(40, activation='relu',  use_bias=True)(x)

output_layer = Dense(nl, activation='linear', use_bias=True)(x)

model = Model(input_layer, output_layer)
        
adam = optimizers.Adam(lr=lr, beta_1=0.9, beta_2=0.999, epsilon=None, decay=0.0, amsgrad=False)

model.compile(loss='mse', optimizer=adam, metrics=[cod])
\end{lstlisting}

\begin{lstlisting}[frame=single,language=Python, caption=Sample code for the CNN used in this study. ,label={lst:cnn}]
inputf = Input(shape=(nx,ny,nci))
        
x = Conv2D(16, (3, 3), activation='relu', padding='same')(inputf)
x = Conv2D(8, (3, 3), activation='relu', padding='same')(x)
x = Conv2D(8, (3, 3), activation='relu', padding='same')(x)
x = Conv2D(8, (3, 3), activation='relu', padding='same')(x)
x = Conv2D(8, (3, 3), activation='relu', padding='same')(x)
x = Conv2D(16, (3, 3), activation='relu', padding='same')(x)

output = Conv2D(nco, (3, 3), activation='linear', padding='same')(x)

model = Model(inputf, output)

adam = optimizers.Adam(lr=lr, beta_1=0.9, beta_2=0.999, epsilon=None, decay=0.0, amsgrad=False)

model.compile(loss='mse', optimizer=adam, metrics=[cod])
\end{lstlisting}

\bibliographystyle{unsrt} 
\bibliography{ref}   

\end{document}